\documentclass[aps, prd, amsmath, floats, floatfix, twocolumn, nofootinbib, superscriptaddress]{revtex4-1}
\usepackage{graphicx}
\usepackage{color}
\usepackage{latexsym}

\begin{document}

\title{Accretion disks around the Gibbons-Maeda-Garfinkle-Horowitz-Strominger charged black holes}

\author{R.Kh. Karimov}
\affiliation{Zel'dovich International Center for Astrophysics, Bashkir State Pedagogical University, 3A, October Revolution Street, Ufa 450008, RB, Russia}

\author{R.N. Izmailov}
\email[Corresponding author: ]{izmailov.ramil@gmail.com}
\affiliation{Zel'dovich International Center for Astrophysics, Bashkir State Pedagogical University, 3A, October Revolution Street, Ufa 450008, RB, Russia}

\author{A. Bhattacharya}
\affiliation{Department of Mathematics, Kidderpore College, 2, Pitamber Sircar Lane, Kolkata 700 023, WB, India}

\author{K.K. Nandi}
\affiliation{Zel'dovich International Center for Astrophysics, Bashkir State Pedagogical University, 3A, October Revolution Street, Ufa 450008, RB, Russia}
\affiliation{High Energy Cosmic Ray Research Center, University of North Bengal, Darjeeling 734 013, WB, India}


\begin{abstract}
It seems surprising that the emissivity properties of the accretion disk (\textit{\`{a} la} Page and Thorne) surrounding the Gibbons-Maeda-Garfinkle-Horowitz-Stro- minger (GMGHS) black holes of heterotic string theory have not yet been studied. To fill this gap in the literature, we study the emissivity properties of the thin accretion disks around these black holes both in the Einstein and in the string frame using the Page-Thorne model. For illustration, we choose as a toy model a stellar-sized spherically symmetric black hole and find that, while the emissivity properties do not significantly differ from those of Reissner-Nordstr\"{o}m and Schwarzschild black holes, they remarkably differ at GMGHS extreme limits corresponding to naked singularity and wormhole at higher frequencies. These differences provide a novel way to speculatively conclude about different types of objects from the observational point of view.
\end{abstract}

\maketitle


\section{Introduction}

String theory is a promising candidate for a consistent quantum theory of gravity and needless to say that the characteristics of black holes (BH) in string theory would be utmost interest. The predictions of string theory differ from those of general relativity and one of the reasons for this difference is the presence of a scalar field called dilaton that can change the properties of the BH geometries.

The spherically symmetric static charged BH solution in low energy heterotic string theory in four dimension was found by Gibbons and Maeda~[1] and independently by Garfinkle, Horowitz, Strominger~[2], which from now on will be referred to as the Gibbons-Maeda-Garfinkle-Horowitz-Strominger (GMGHS) solution. These works generated enormous interest in the dilatonic charged BHs (see, e.g., [3-16] and references therein). In particular, the GMGHS BH spacetime can be described either in the Einstein frame (EF) or in the conformally related string frame (SF). (Unless the frame is specifically mentioned, the solution will be understood to be in EF). In EF, the action is in the form of the Einstein-Hilbert action, while in the SF strings directly couple to the metric as $e^{2\phi }g_{\mu \nu }$, where $\phi $ is the dilaton and $g_{\mu \nu }$ is the EF metric. Even though the solutions in the two frames are related by a conformal transformation so that they are mathematically isomorphic to each other~[3], there are differences in some of the physical properties of the BH solutions in these two frames~[4]. For instance, Sagnac delay for rotating Kerr-Sen metric of heterotic string theory was studied and an estimate of assumed terrestrial dilatonic charge was given in [5]. Strong gravitational lensing by charged stringy BHs do not produce any significant string effect on the Schwarzshild BH as was shown by Bhadra [6]. Timelike geodesics of particles around GMGHS BH were investigated in [7]. The magnetically charged GMGHS interior spacetime was studied in [8]. Particle collision near the horizon of GMGHS BH was studied in [9]. Quasinormal mode frequencies in the string BH were evaluated by using WKB approximation with P\"{o}schl-Teller potential in [10]. Some notable works on the implications of electric charge on various relativistic observables in the context of general relativistic BHs can be found in [17-21].

In addition to the above studies, which are by no means exhaustive, the accretion disk properties could be yet another diagnostic for distinguishing various types of objects. The first comprehensive study of accretion disks using a Newtonian approach was made in [22]. Later a general relativistic model of thin accretion disk was developed in three seminal papers, by Novikov and Thorne [23], Page and Thorne [24] and Thorne [25] under the assumption that the disk is in a steady-state, that is, the mass accretion rate $\dot{M}_{0}$ is constant in time and does \textit{not} depend of the radius of the disk. The disk is further supposed to be in hydrodynamic and thermodynamic equilibrium, which ensure a black body electromagnetic spectrum and properties of emitted radiation. The thin accretion disk model further assumes that individual particles are moving on Keplerian orbits, but for this to be true the central object should have weak magnetic field, otherwise the orbits in the inner edge of the disk will be deformed. More recently, the properties of the accretion disk around exotic central objects, such as wormholes (WH) [26,27], and non-rotating or rotating quark, boson or fermion stars, brane-world BHs, gravastars or naked singularities (NS) [28-44], $f(R)$ modified gravity models [44-46] have been studied. One of the most promising method to distinguish the types of astrophysical objects and their spin through their accretion disk properties is the profile analysis of iron line for different spacetimes [47-52]. While various accreting objects have been considered, the study of emissivity properties of the GMGHS objects, for which the celebrated Page-Thorne model is the most ideal one, have somehow been left out in the literature, to our knowledge.

One recent work by Bahamonde and Jamil [11] pertains to fluid motion (as opposed to particle motion) in different spherically symmetric spacetimes, where the authors focused on the radial variation of the fluid velocity, the density and the accretion rate\ $\dot{M}_{0}$ of the fluid to the GMGHS BH. The work in [11], though useful in its own right, is distinct from the Page-Thorne emission model since it studied only the non-emissivity aspects of the fluid flow with the predicted mass accretion rate $\dot{M}_{0}$ depending on the radius of the disk. Also the critical radius $r_{\textmd{\scriptsize{c}}}$ used in [11] is not the marginally stable radius $r_{\textmd{\scriptsize{ms}}}$ required in the Page-Thorne model. On the other hand, we are motivated by the understanding that genuine observable signatures of the accretion disk should be provided by an analysis of the properties of radiation emerging from the surface of the disk for which the Page-Thorne model is most suitable.

The present paper is thus devoted to studying the kinematic and emissivity properties of a central object represented by stringy GMGHS solutions (not necessarily BHs) in the EF and SF using the Page-Thorne model. We shall analyze the luminosity spectra, flux of radiation, temperature profile, efficiency etc. In particular, we wish to see how the kinematic and accretion features compare between EF and SF including their extreme counterparts and with similar features from Reissner-Nordstr\"{o}m and Schwarzschild BH of general relativity. We shall assume for numerical illustration a toy model of a central object with mass $15M_{\odot }$ and accretion rate $\dot{M}_{0}=10^{18}$ gm.sec$^{-1}$, which could be a BH, a WH or a NS.

The paper is organized as follows: In Sec.2, we give a brief preview of the GMGHS solutions and in Sec.3, summarize the main formulas relating to the accretion phenomenon to be used in the paper. In Sec.4, we present the kinematic and accretion properties in two frames and compare how they differ from those for the Reissner-Nordstr\"{o}m and Schwarzschild BH. In Sec.5, the obtained results are summarized. We take units such that $8\pi G=c=1$, metric signature ($-,+,+,+$) and greek indices run from $0$ to $3$.

\section{GMGHS solutions}

In this section, a brief preview of the action and the static spherically symmetric dilaton BHs are given. In the EF, the GMGHS action is [1,2]
\begin{equation}
S_{\textmd{\scriptsize{EF}}}=\int d^{4}x\sqrt{-g}\left[ R_{(g)}-2(\bigtriangledown \phi
)^{2}-e^{-2\phi }F_{\mu \nu }F^{\mu \nu }\right],
\end{equation}%
where $\phi $ is a dilaton, $R_{(g)}$ is the scalar curvature related to $g_{\mu \nu }$, and $F_{\mu \nu }$ is the Maxwell field. The line element
representing a 4-dimensional charged dilatonic GMGHS BH in the EF is given by
\begin{eqnarray}
d\tau _{\textmd{\scriptsize{Mag,EF}}}^{2}&=&-\left( 1-\frac{2M}{r}\right) dt^{2}+\left( 1-\frac{2M}{r}\right)^{-1}dr^{2} \nonumber\\
&&+r\left( r-\frac{Q^{2}}{M}\right) (d\theta^{2}+ \sin^{2}{\theta} d\varphi^{2}),
\end{eqnarray}%
where $M$ is the mass and $Q$ is the magnetic charge. The Maxwell field is given by
\begin{equation}
F=Q\sin {\theta }d\theta \wedge d\varphi
\end{equation}%
and the dilaton field $\phi $ is defined as
\begin{equation}
e^{-2\phi }=e^{-2\phi _{0}}\left( 1-\frac{Q^{2}}{Mr}\right) ,
\end{equation}%
where $\phi _{0}$ is the asymptotic value of the dilaton. As we consider only asymptotically flat cases, we will assume $\phi _{0}\equiv 0$. Solution (2) describes BHs of mass $M$ and charge $Q$ when $Q/M$ is sufficiently small. For $Q=\sqrt{2}M$, the event horizon $r=2M$ becomes singular. (Figs.1,4,5 show the properties of this singularity at $r=2M$: potential $V_{\textmd{\scriptsize{eff}}}$, Flux of radiation and Temperature diverge).

Since the metric, for fixed $\theta $ and $\varphi $, is the same as that of Schwarzschild, $r=2M$ is a regular event horizon only when $Q<\sqrt{2}M$. Note that the area goes to zero at $r=Q^{2}/M$ $<2M$ causing this surface to be singular since the Kretschmann scalar and the dilaton $\phi $ diverge there. However, the surface is hidden under $r=2M$ and no information can emerge from it to outside observers (All observable quantities also diverge on that surface). The dilatonic charge for the charged BH (2) is
\begin{equation}
D=-\frac{Q^{2}}{2M},
\end{equation}%
where $D$ is not a new free parameter in (2) since once the asymptotic value of $\phi $ is fixed, it is determined by $M$ and $Q$, and is always negative. The dilaton charge is also responsible for a long-range, attractive force between BHs.

Electrically charged solutions may be obtained by a duality rotation defined by
\begin{equation}
\tilde{F}_{\mu \nu }=\frac{1}{2}e^{-2\phi }\epsilon _{\mu \nu }^{\lambda
\rho }F_{\lambda \rho }.
\end{equation}%
The equations of motion (2) are invariant under $F\rightarrow \tilde{F}$ and $\phi \rightarrow -\phi $. Such solutions can therefore be obtained by simply changing the sign of $\phi $ while keeping the metric fixed. This implies that the dilaton charge is $D$ is positive for electrically charged BHs in the EF.

The effective action and equations of motion can be further modified by applying the conformal transformation $\widetilde{g}_{\mu \nu }= e^{2\phi}g_{\mu\nu }$ thus obtaining the action in the SF
\begin{equation}
S_{\textmd{\scriptsize{SF}}}=\int d^{4}x\sqrt{-\widetilde{g}}e^{-2\phi }\left[ R_{(\widetilde{g})}-4(\bigtriangledown \phi )^{2}- F_{\mu\nu}F^{\mu\nu}\right],
\end{equation}%
in which the space-time coordinates have been left unchanged and $R_{(\widetilde{g})}$ is the Ricci curvature from $\widetilde{g}_{\mu \nu }$. Upon transforming to SF one obtains magnetically charged GMGHS BH metric is given by [15,16]:
\begin{eqnarray}
d\tau _{\textmd{\scriptsize{Mag,SF}}}^{2}&=&-\frac{\left( 1-\frac{2M}{r}\right) }{\left( 1-\frac{Q^{2}}{Mr}\right) }dt^{2}+\frac{dr^{2}}{\left(1- \frac{2M}{r}\right)\left(1-\frac{Q^{2}}{Mr}\right)} \nonumber\\
&&+r^{2}(d\theta ^{2}+\sin^{2}{\theta} d\varphi ^{2}).
\end{eqnarray}%

This is the metric that appears in the string $\sigma $ model. The Kretschmann scalar diverges at $r=Q^{2}/M$, hence it is a singular surface. For SF, the statement that the horizon is singular when $Q^{2}=2M^{2}$ is actually irrelevant. This is because strings do not couple to the metric $g_{\mu \nu }$ but rather to $e^{2\phi }g_{\mu \nu }$. For $Q^{2}<2M^{2}$, this again describes a BH with an event horizon at $r_{\textmd{\scriptsize{eh}}}=2M$. We have simply rescaled the metric by a conformal factor, which is finite everywhere outside (and on) the horizon. However, at the extremal value $Q^{2}=2M^{2}$, the metric becomes
\begin{eqnarray}
d\tau _{\textmd{\scriptsize{WH,SF}}}^{2}&=&-dt^{2}+\left( 1-2M/r\right)^{-2}dr^{2} \nonumber\\
&&+r^{2}(d\theta ^{2}+\sin^{2}{\theta} d\varphi ^{2}).
\end{eqnarray}%
The geometry of a $t=$ const. surface in this spacetime is identical to that
of a static slice in the extreme Reissner-Nordstr\"{o}m metric. But the
horizon, along with the singularity inside it, have completely disappeared.
In its place have appeared a WH. This metric, with $r>2M$, is globally
static and geodesically complete and has all the properties of a traversable
Morris-Thorne WH with a regular throat at $r_{\textmd{\scriptsize{th}}}=2M$ with redshift
function $\Phi =0$ and a shape function $b(r)=4M\left( 1-\frac{M}{r}\right) $
having an imbedding surface%
\begin{eqnarray}
z(r)&=&4\sqrt{M}\left[ \sqrt{r-M}\right. \nonumber\\
&&\left.-\sqrt{M}\textmd{arctanh}\left\{ \sqrt{\frac{r}{%
M}-1}\right\} \right].
\end{eqnarray}%

The solution of the electrically charged GMGHS solution in the SF is given
by [15,16]
\begin{eqnarray}
d\tau _{\textmd{\scriptsize{Elec,SF}}}^{2}&=&-\frac{\left( 1+\frac{Q^{2}-2M^{2}}{Mr}\right) }{%
\left( 1+\frac{Q^{2}}{Mr}\right) ^{2}}dt^{2}+\frac{dr^{2}}{\left( 1+\frac{Q^{2}-2M^{2}}{Mr}\right) } \nonumber\\
&&+r^{2}(d\theta ^{2}+\sin^{2}{\theta} d\varphi ^{2}).
\end{eqnarray}%
This solution describes a BH, when $Q^{2}<2M^{2}$, but in the extremal case (%
$Q^{2}=2M^{2}$) the resulting solution is
\begin{eqnarray}
d\tau _{\textmd{\scriptsize{NS,SF}}}^{2}&=&-\left( 1+2M/r\right)
^{-2}dt^{2}+dr^{2} \nonumber\\
&&+r^{2}(d\theta ^{2}+\sin^{2}{\theta} d\varphi ^{2}).
\end{eqnarray}%
which describes a singularity at $r=0$ since the Kretschmann scalar diverges
there but this divergence is not covered by an event horizon, and so the
point $r=0$ represents a NS.

\begin{figure*}
\begin{center}
\includegraphics[type=pdf,ext=.pdf,read=.pdf,width=8.5cm]{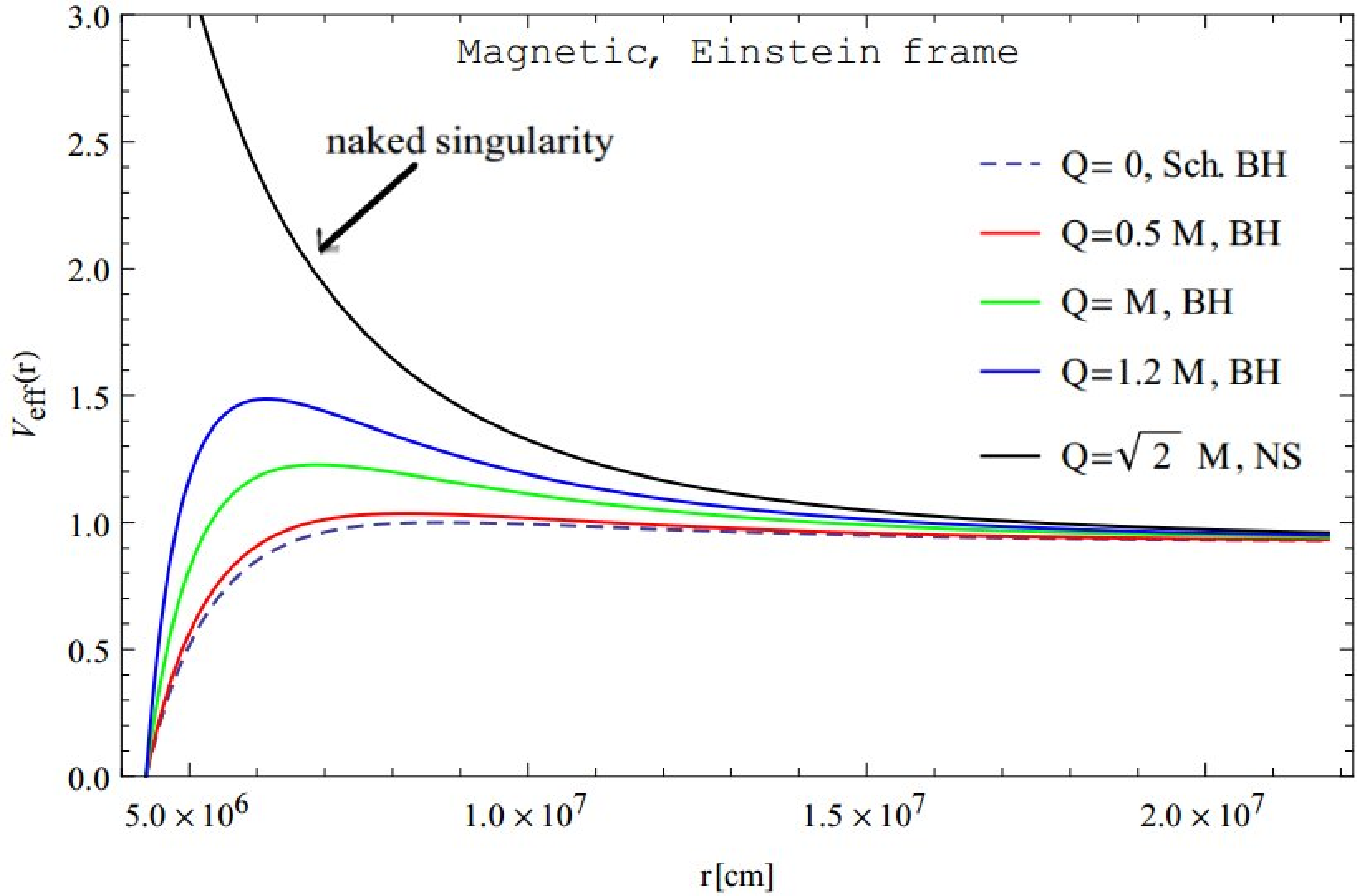}
\hspace{0.3cm}
\includegraphics[type=pdf,ext=.pdf,read=.pdf,width=8.5cm]{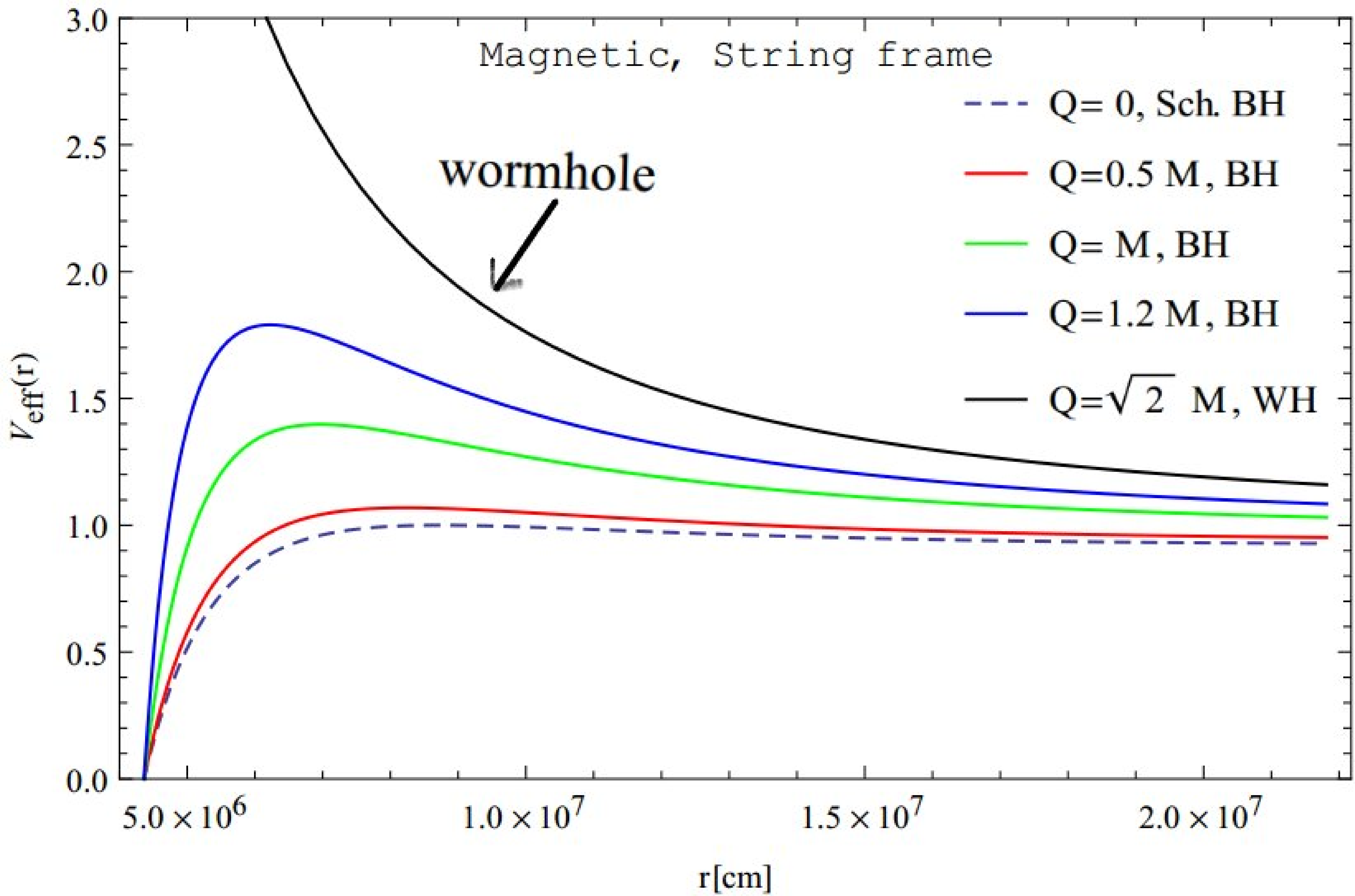} \\
\vspace{0.5cm}
\includegraphics[type=pdf,ext=.pdf,read=.pdf,width=8.5cm]{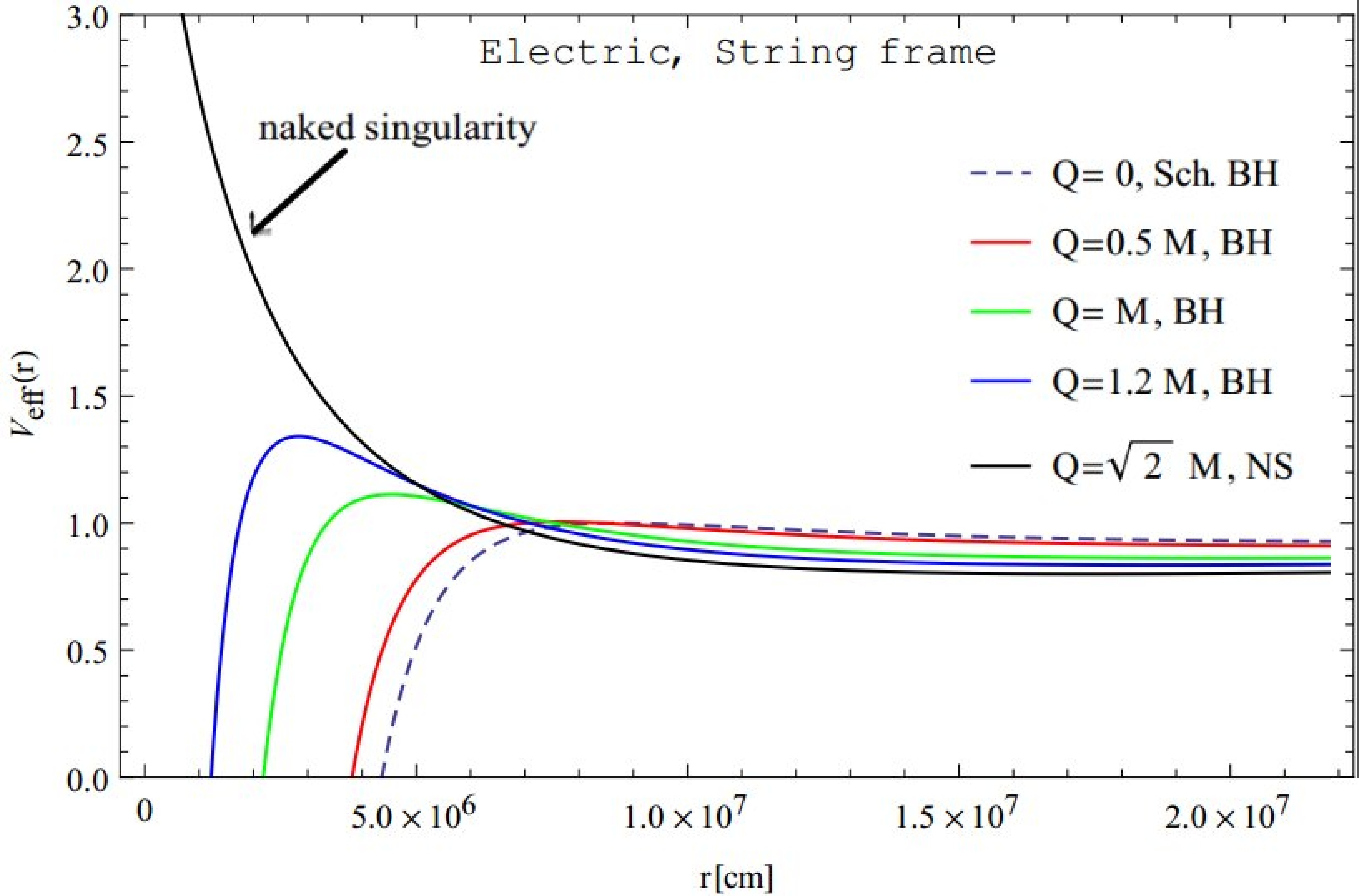}
\hspace{0.3cm}
\includegraphics[type=pdf,ext=.pdf,read=.pdf,width=8.5cm]{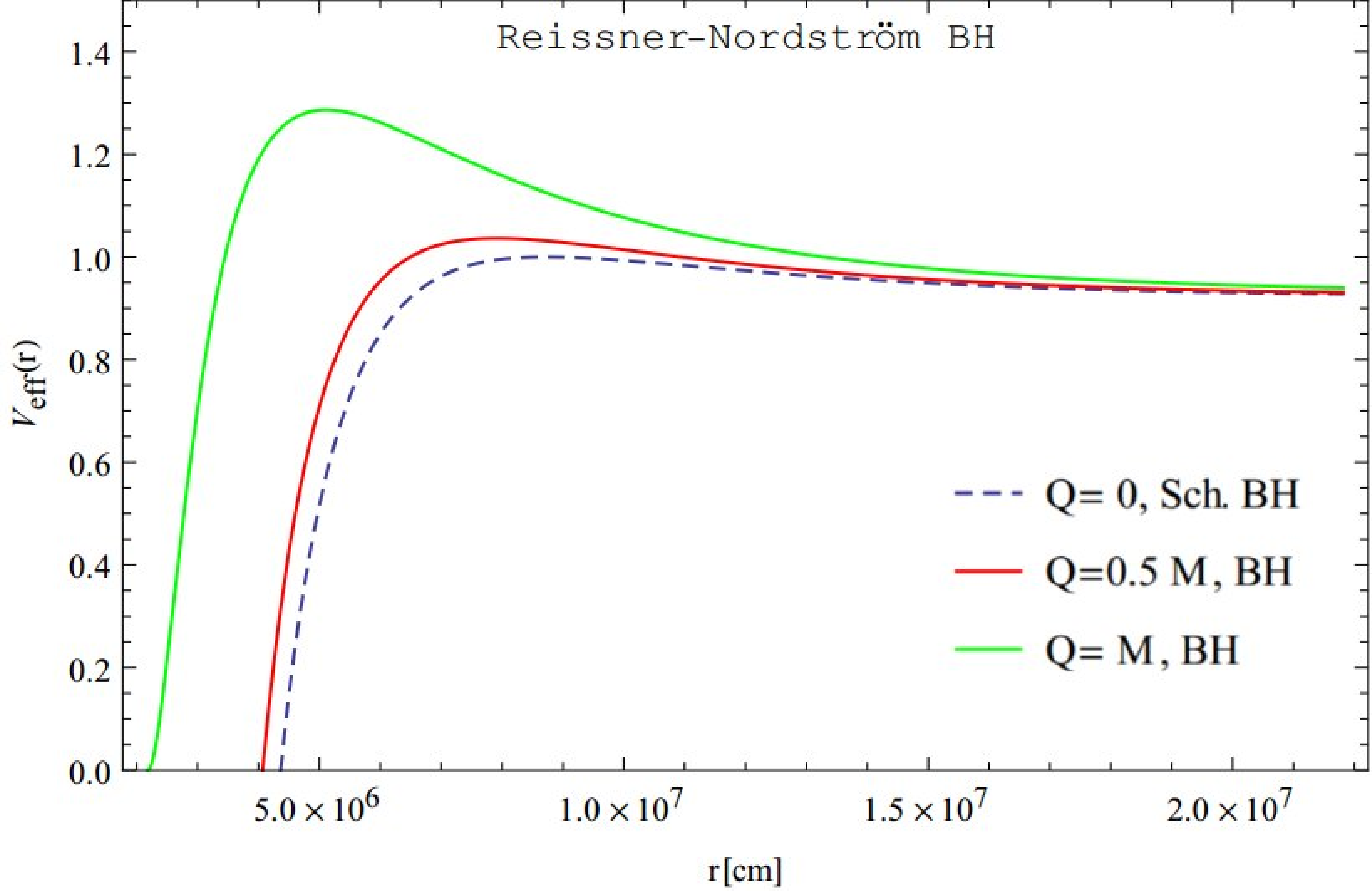}
\end{center}
\caption{The effective potential $V_{\textmd{\scriptsize{eff}}}(r)$ for a GMGHS BH in the EF (top left hand), magnetically charged one in SF (top right hand), electrically charged one in the SF (bottom left hand) and Reissner-Nordstr\"{o}m BH (bottom right hand). The specific angular momentum of the orbiting particle is chosen to be $\widetilde{L}=4M$. The potentials for $Q<\protect\sqrt{2}M$ do not show appreciable difference with those of the Reissner-Nordstr\"{o}m and Schwarzschild BH (Fig.1d). At the extreme limit, $Q=\protect\sqrt{2}M$, the potential diverges at the NS radii (Figs.1a,c) and at the WH throat (Fig.1b). The potential coincides with that of the Schwarzschild asymptotically.}
\label{Veff}
\end{figure*}

\begin{figure*}
\begin{center}
\includegraphics[type=pdf,ext=.pdf,read=.pdf,width=8.5cm]{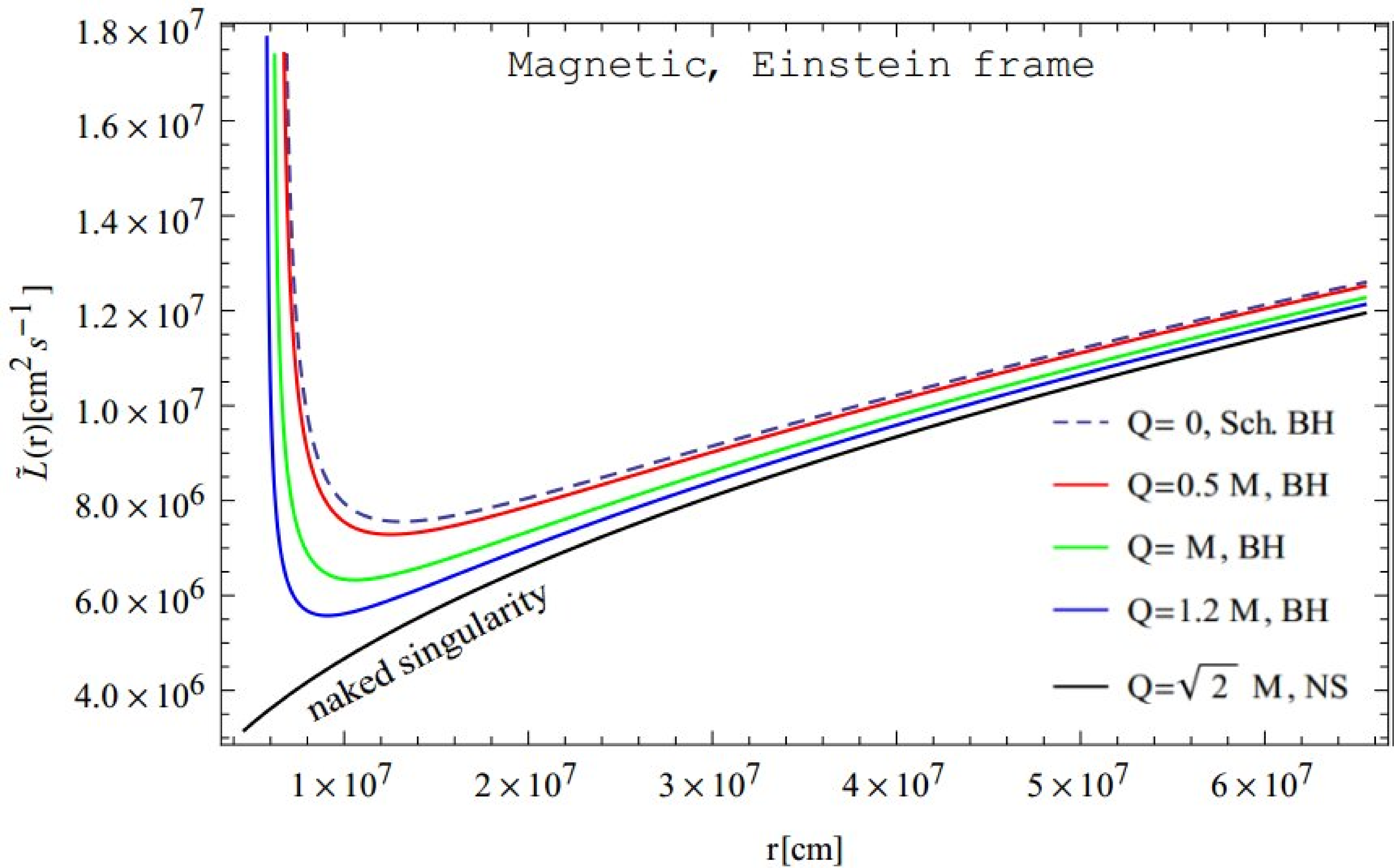}
\hspace{0.3cm}
\includegraphics[type=pdf,ext=.pdf,read=.pdf,width=8.5cm]{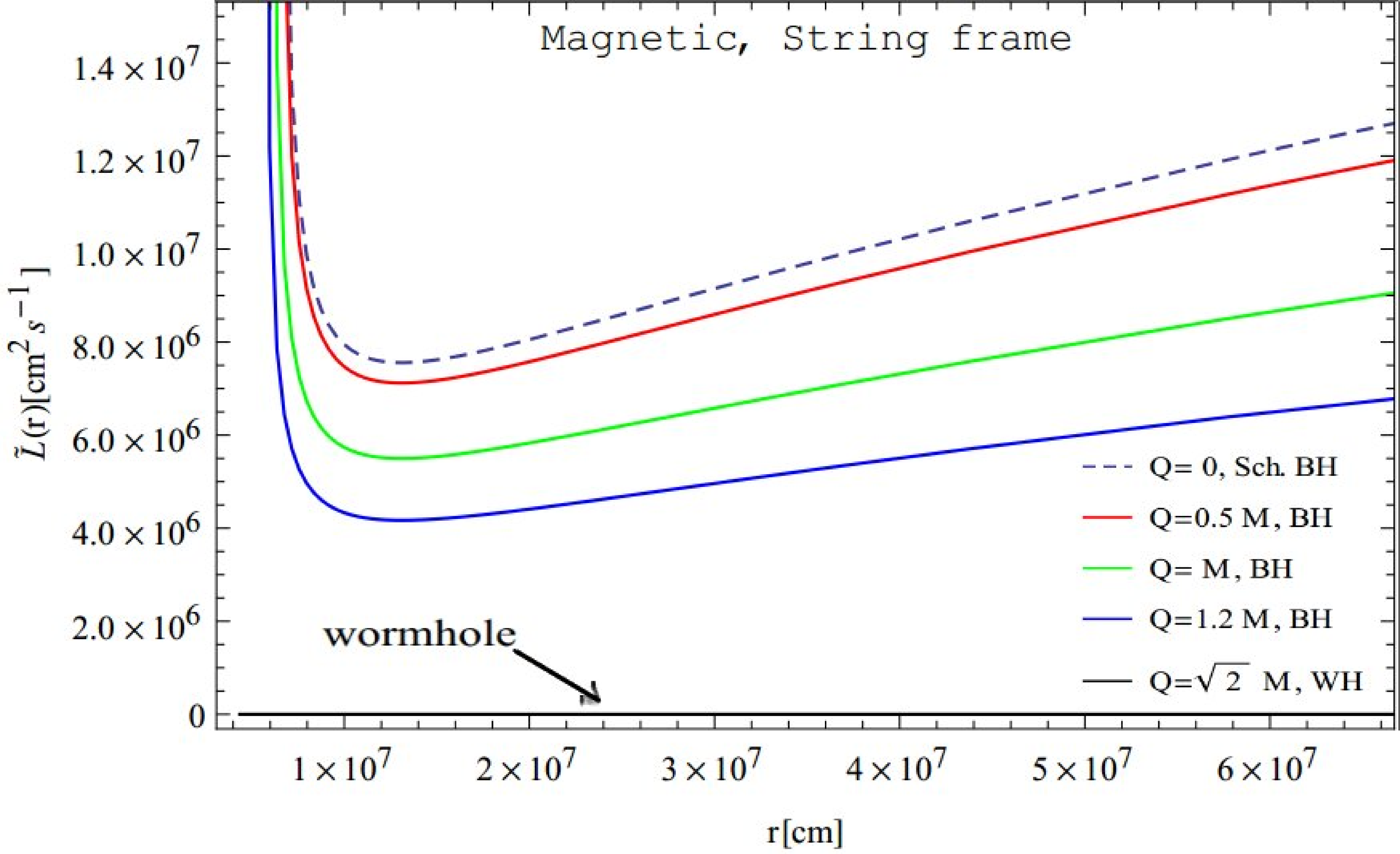} \\
\vspace{0.5cm}
\includegraphics[type=pdf,ext=.pdf,read=.pdf,width=8.5cm]{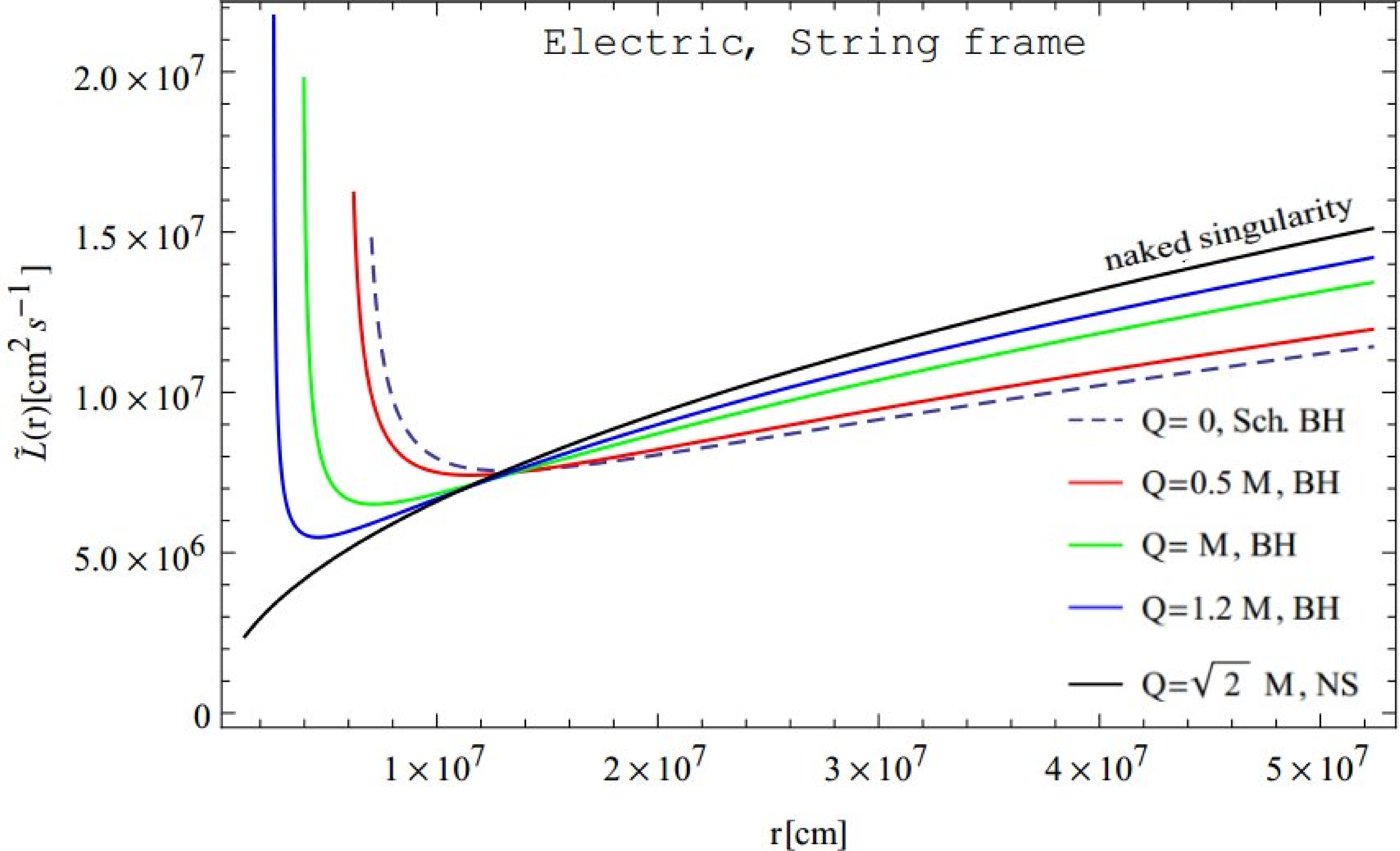}
\hspace{0.3cm}
\includegraphics[type=pdf,ext=.pdf,read=.pdf,width=8.5cm]{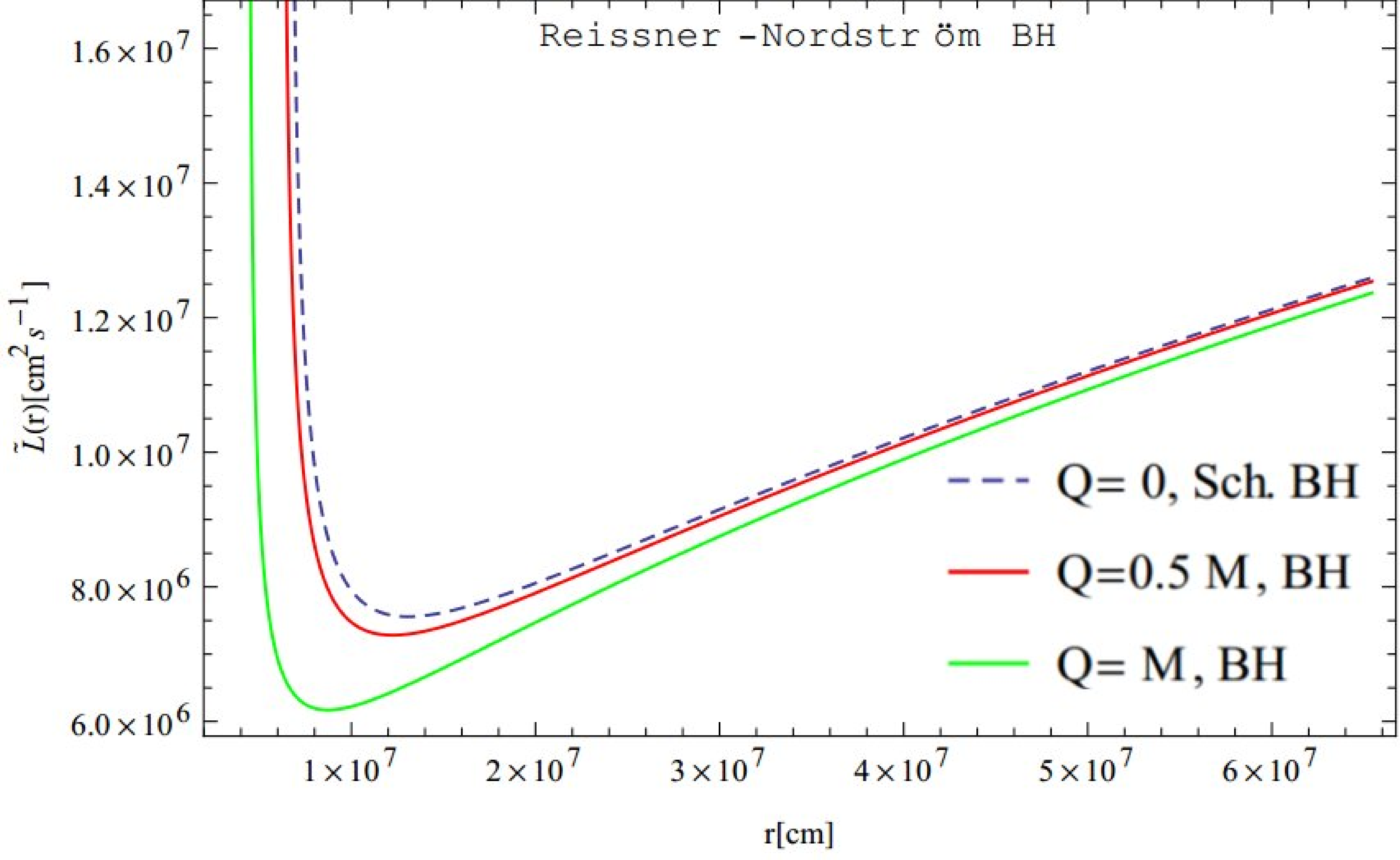}
\end{center}
\caption{The specific angular momentum $\tilde{L}(r)$ of the orbiting particle as a function of the radial coordinate $r$ (in cm) for a GMGHS BH in EF (top left hand), magnetically charged in the SF (top right hand), electrically charged in the SF (bottom left hand) and Reissner-Nordstr\"{o}m BH (bottom right hand) plotted for different values of $Q$. We see that $\tilde{L}\rightarrow 0$ at the NS radii $r\rightarrow 2M$ (Fig.2a), $r\rightarrow 0$ (Fig.2c), but $\tilde{L}\rightarrow 0$ at the WH throat (Fig.2b). Fig.2d displays the behavior only for BHs. All plots show no appreciable difference with those for Schwarzshild BH either in the far field of the accreting object.}
\label{SpAngulMom}
\end{figure*}

\begin{figure*}
\begin{center}
\includegraphics[type=pdf,ext=.pdf,read=.pdf,width=8.5cm]{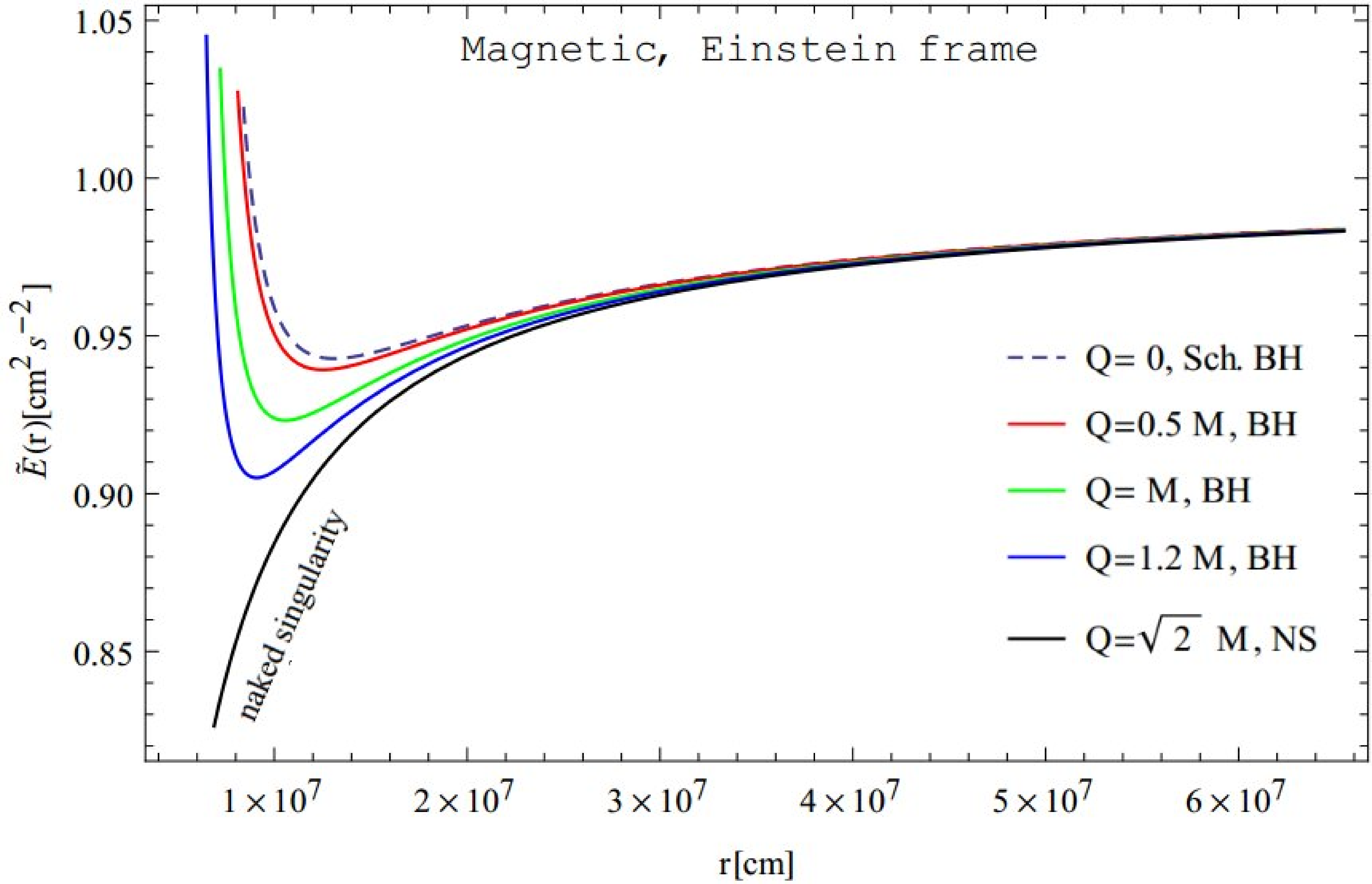}
\hspace{0.3cm}
\includegraphics[type=pdf,ext=.pdf,read=.pdf,width=8.5cm]{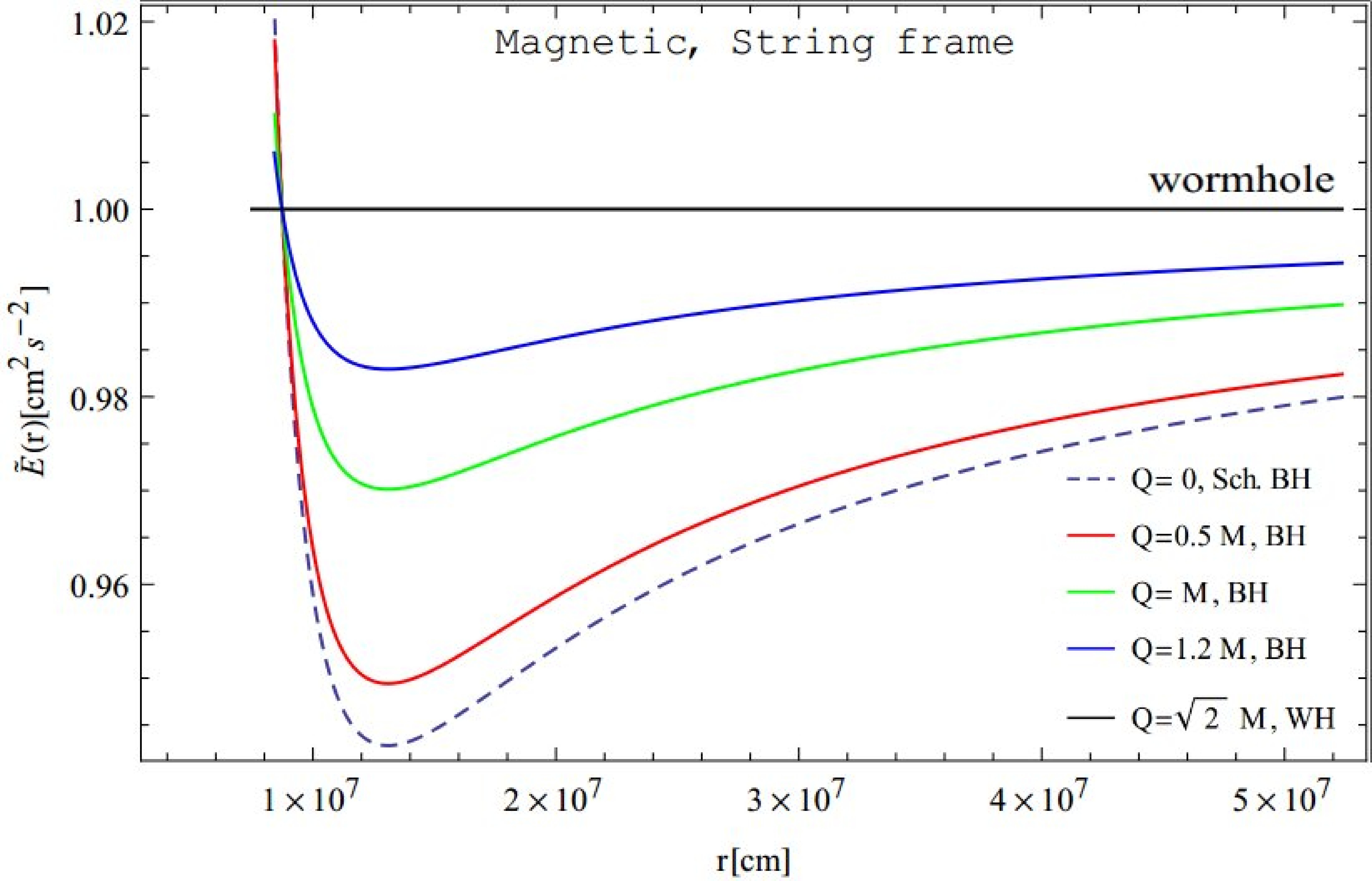} \\
\vspace{0.5cm}
\includegraphics[type=pdf,ext=.pdf,read=.pdf,width=8.5cm]{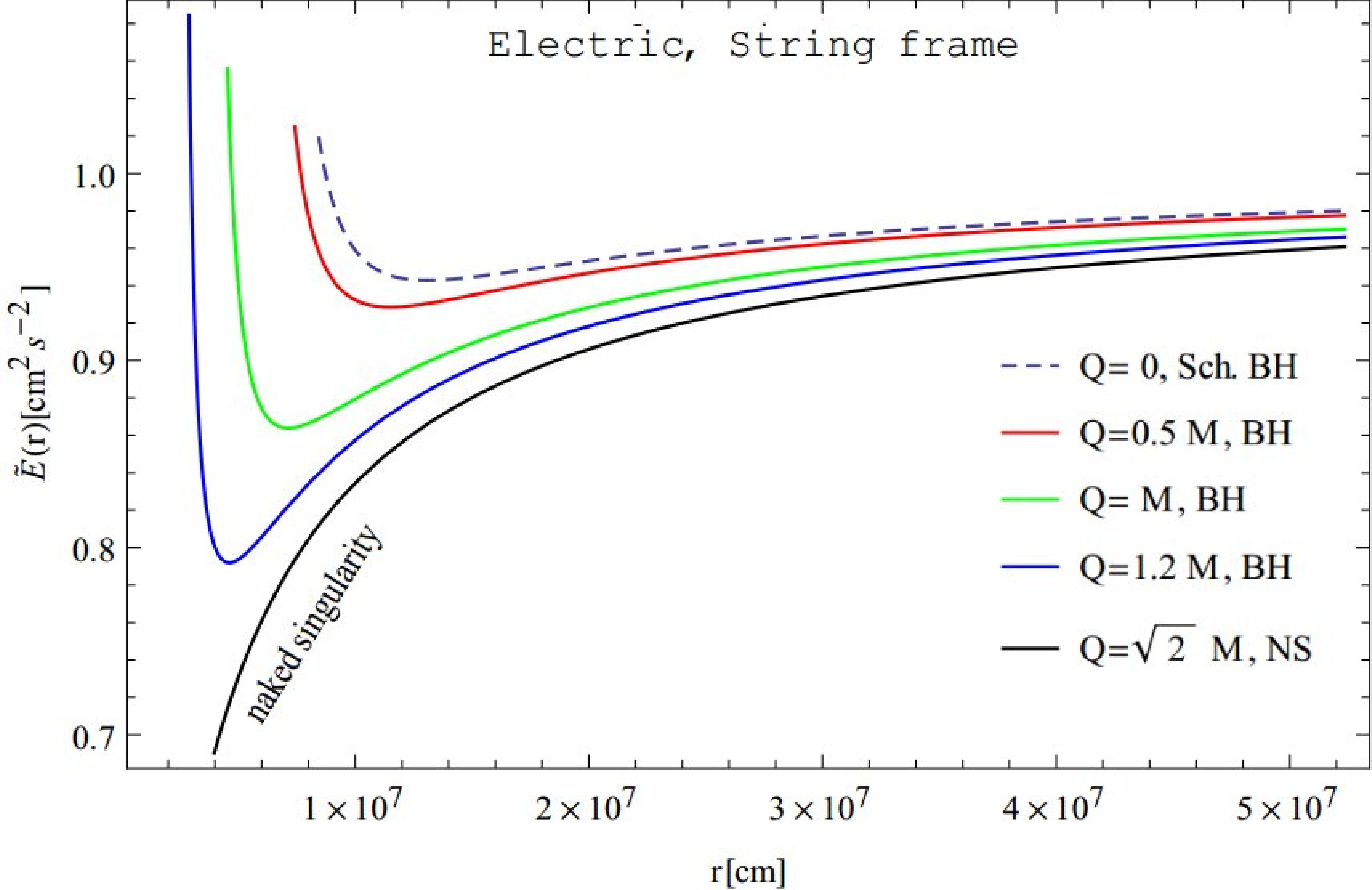}
\hspace{0.3cm}
\includegraphics[type=pdf,ext=.pdf,read=.pdf,width=8.5cm]{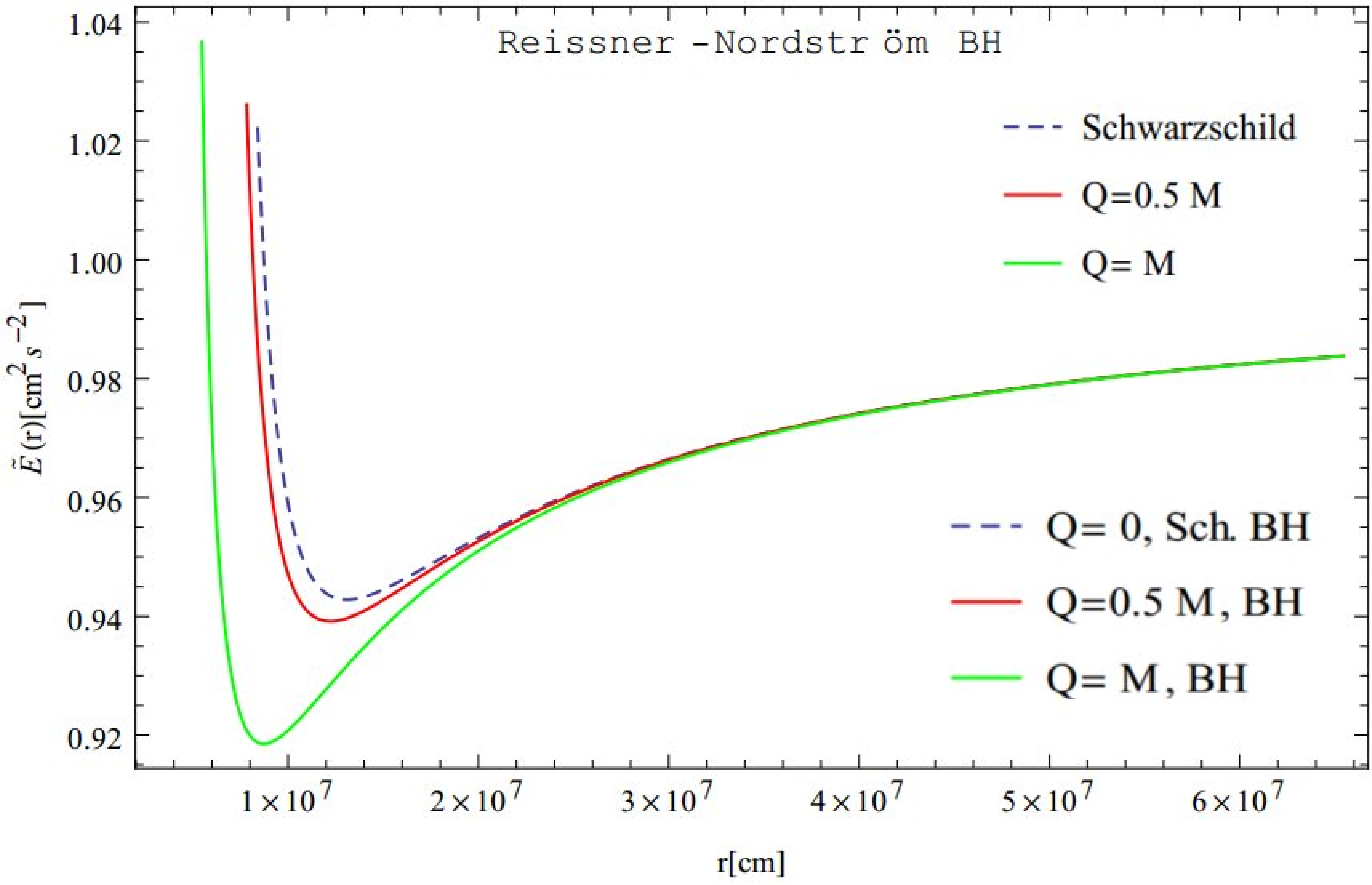}
\end{center}
\caption{The specific energy $\tilde{E}(r)$ of the orbiting particles as a function of the radial coordinate $r$ (in cm) for a GMGHS BH in EF (top left hand), magnetically charged in the SF (top right hand), electrically charged in the SF (bottom left hand) and Reissner-Nordstr\"{o}m BH (bottom right hand) plotted for different values of $Q$. We see that $\tilde{E}\rightarrow 0$ at the NS radii $r\rightarrow 2M$ (Fig.3a), $r\rightarrow 0$ (Fig.3c), but $\tilde{E}$ assumes a constant value at the WH throat (Fig.3b). Fig.3d displays the behavior only for BHs, and there are no appreciable differences in the far field of observation.}
\label{SpEnergy}
\end{figure*}

\begin{figure*}
\begin{center}
\includegraphics[type=pdf,ext=.pdf,read=.pdf,width=8.5cm]{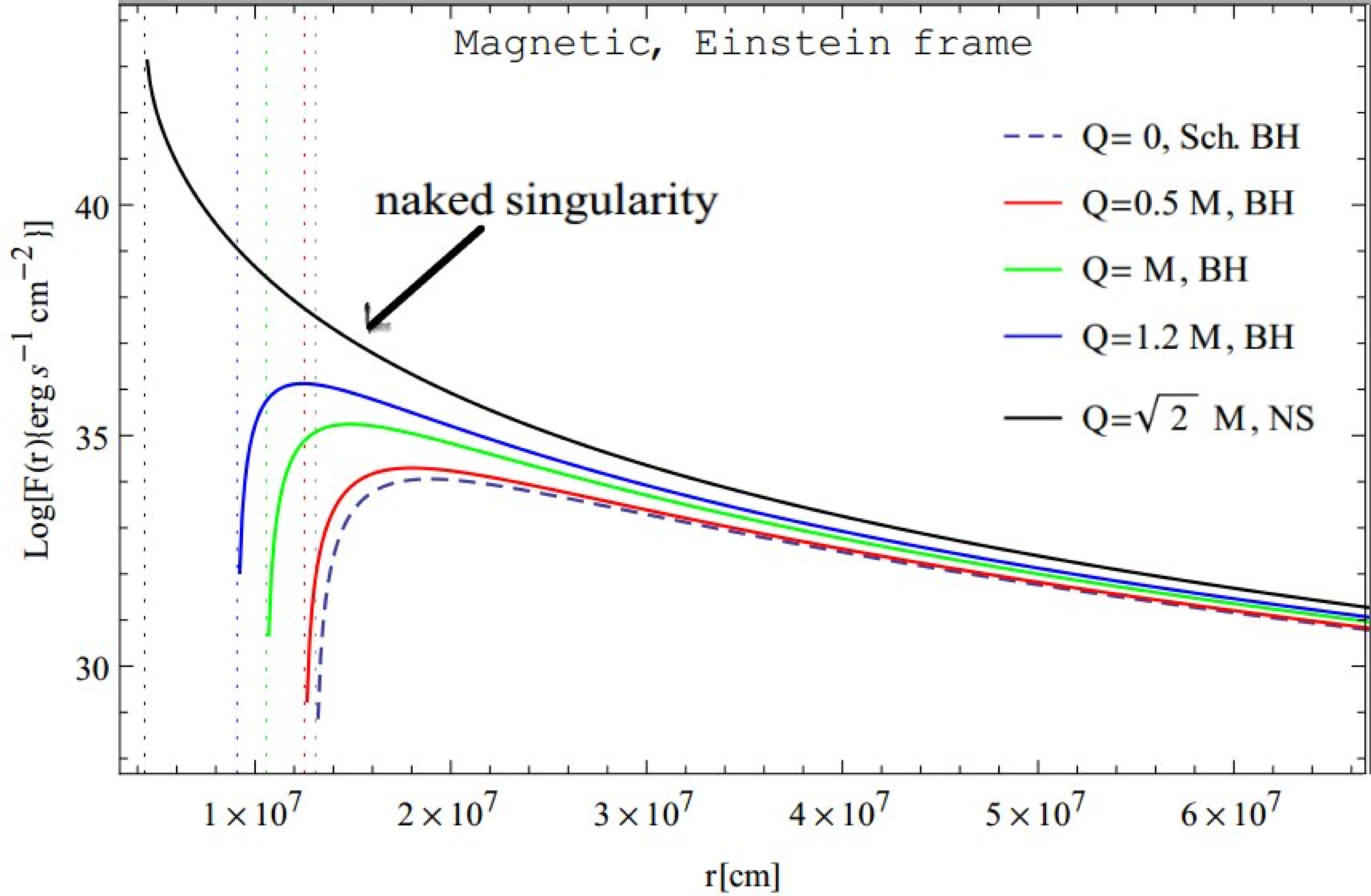}
\hspace{0.3cm}
\includegraphics[type=pdf,ext=.pdf,read=.pdf,width=8.5cm]{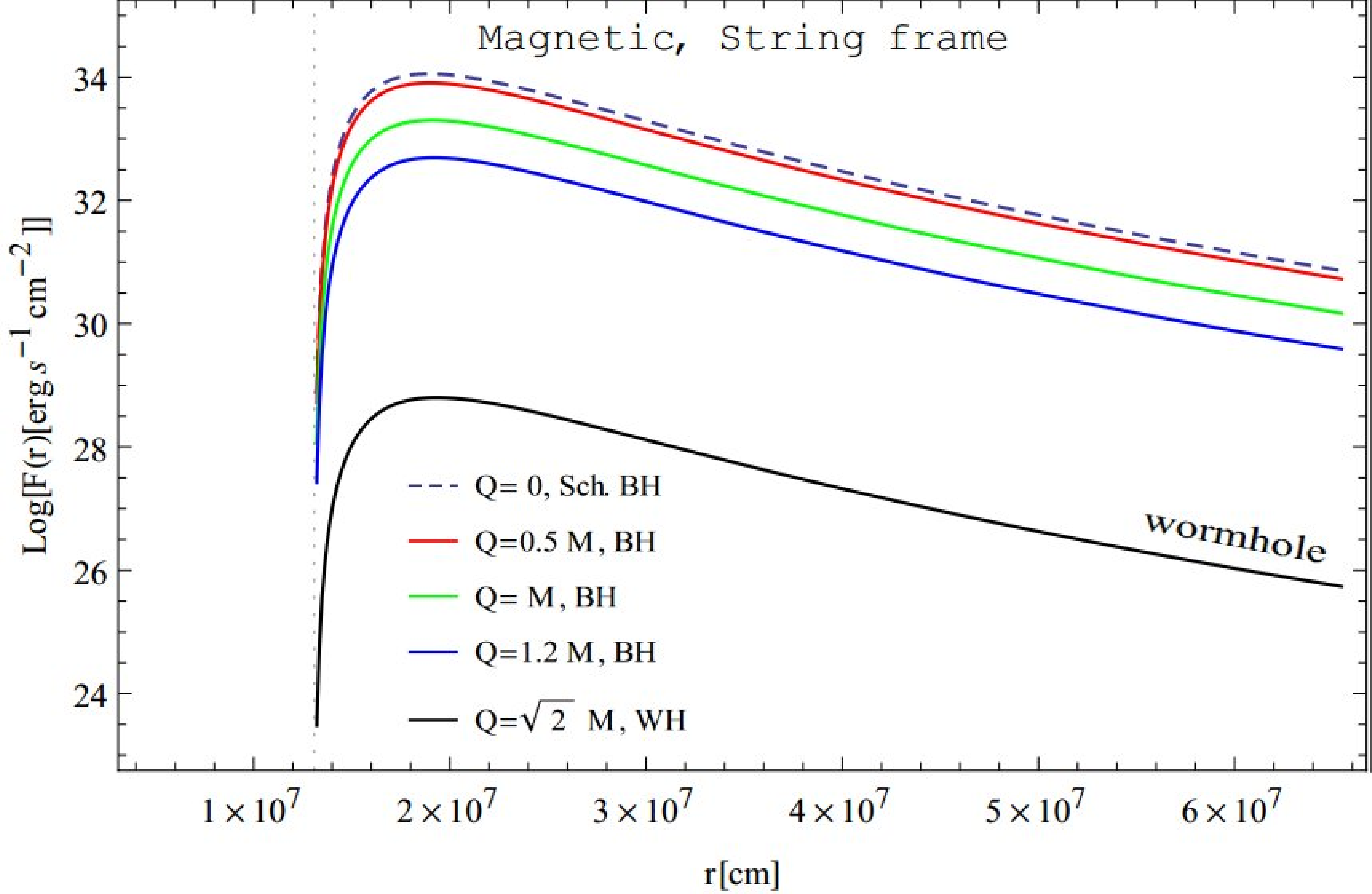} \\
\vspace{0.5cm}
\includegraphics[type=pdf,ext=.pdf,read=.pdf,width=8.5cm]{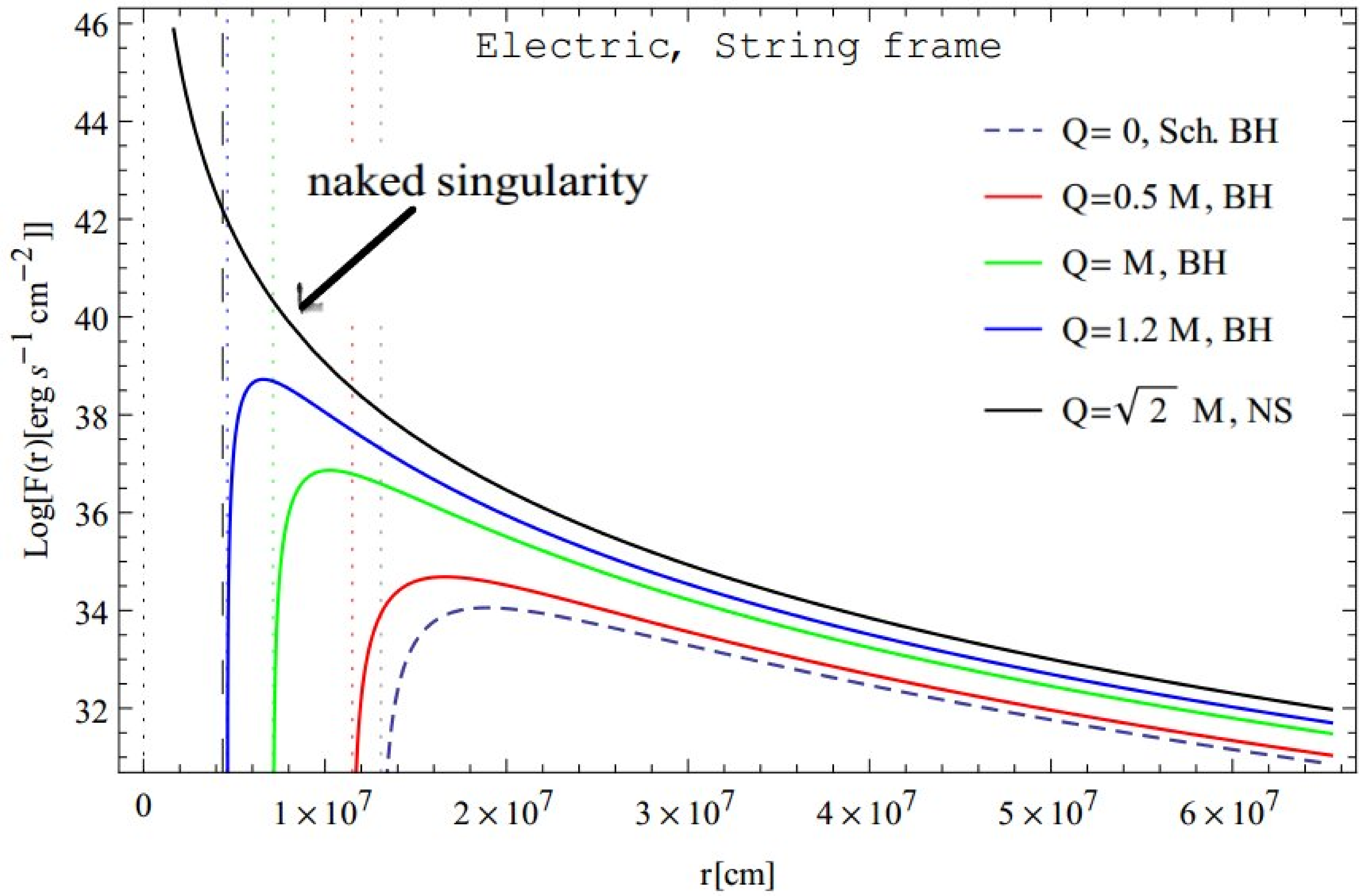}
\hspace{0.3cm}
\includegraphics[type=pdf,ext=.pdf,read=.pdf,width=8.5cm]{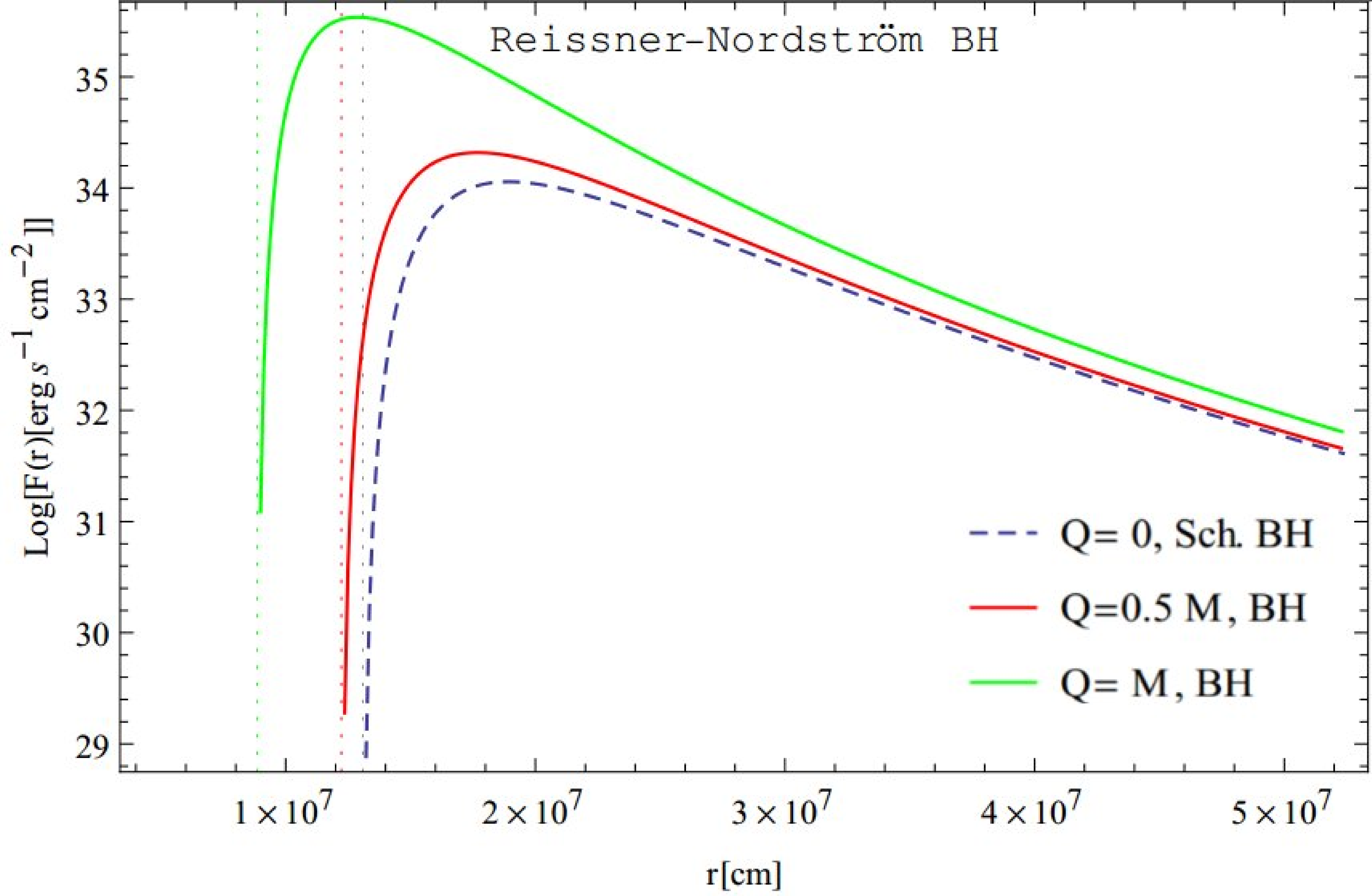}
\end{center}
\caption{The time averaged radiation flux $F(r)$ as a function of the radial coordinate $r$ (in cm) radiated by the disk around a GMGHS BH in EF (top left hand), magnetically charged in the SF (top right hand), electrically charged in the SF (bottom left hand) and Reissner-Nordstr\"{o}m BH (bottom right hand) plotted for different values of $Q$. We see that $F(r)\rightarrow \infty $ at NS radii $2M$ coinciding with $r_{\textmd{\scriptsize{ms}}}$ (Fig.4a), $r=r_{\textmd{\scriptsize{ms}}}\rightarrow 0$ (Fig.4c), but $F(r)$ assumes a finite value at the WH throat (Fig.4b). Fig.4d displays the behavior only for BHs. All plots except that of WH show no appreciable differences at the asymptotic limit.}
\label{Flux}
\end{figure*}

\begin{figure*}
\begin{center}
\includegraphics[type=pdf,ext=.pdf,read=.pdf,width=8.5cm]{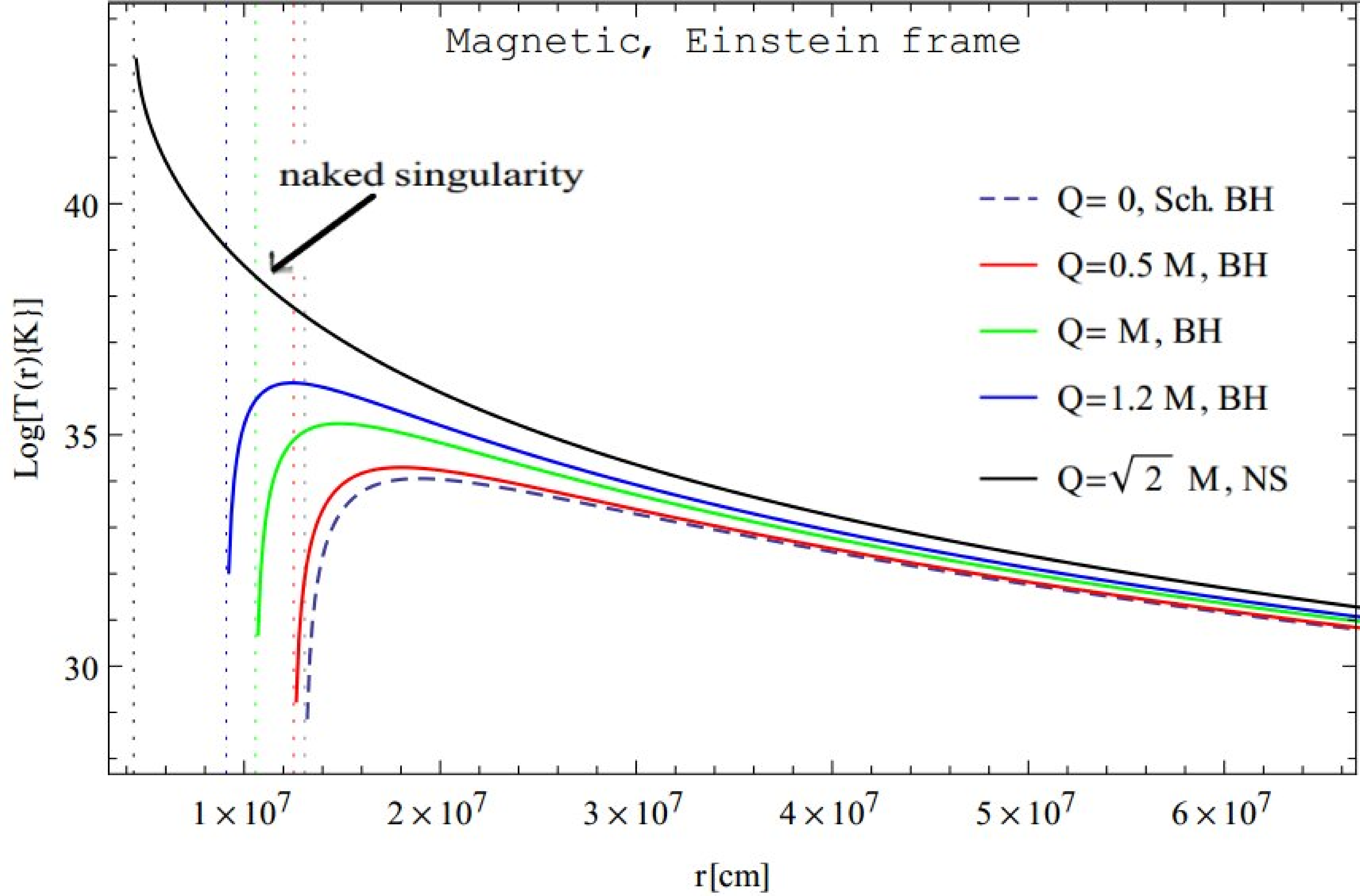}
\hspace{0.3cm}
\includegraphics[type=pdf,ext=.pdf,read=.pdf,width=8.5cm]{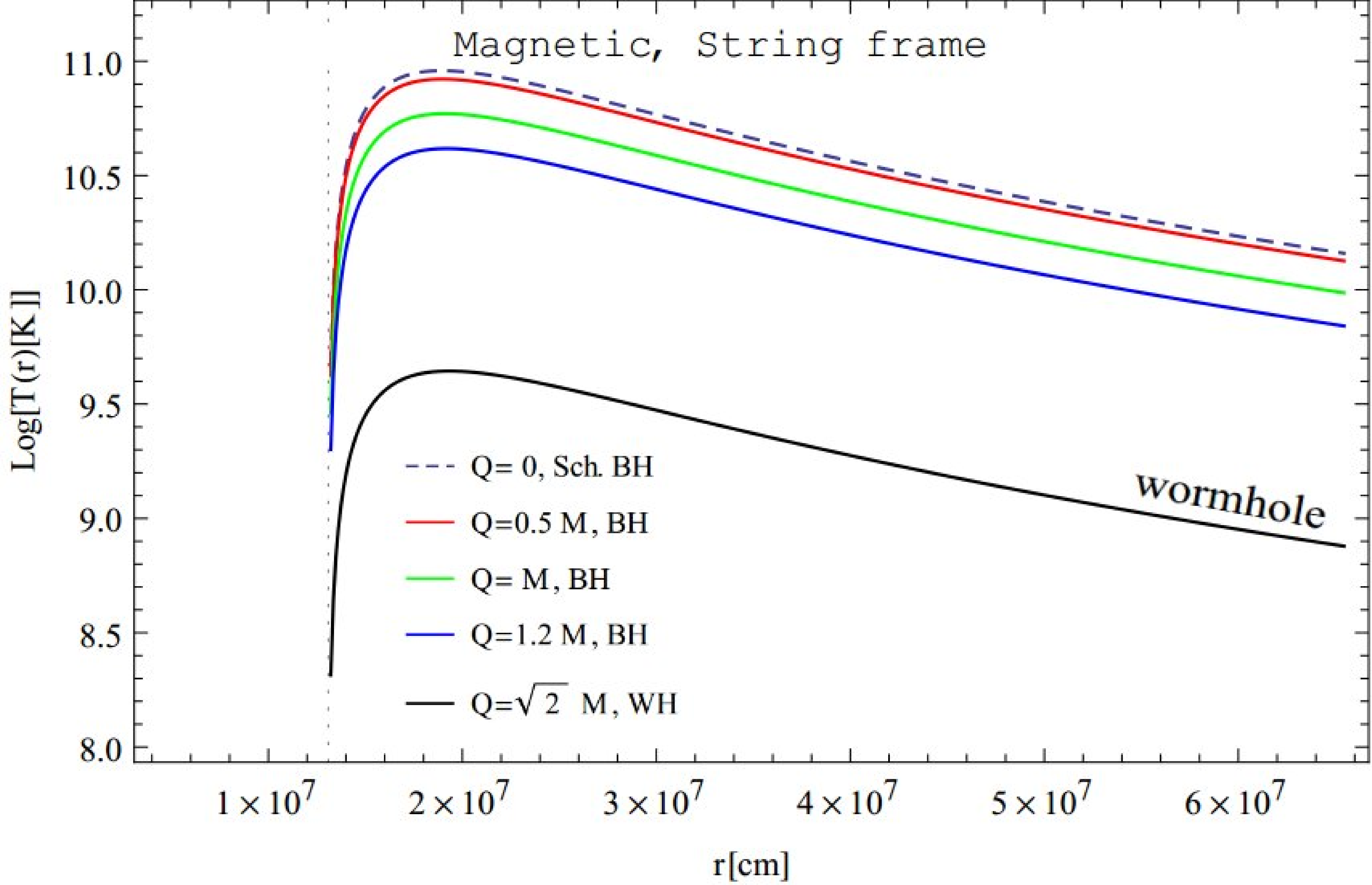} \\
\vspace{0.5cm}
\includegraphics[type=pdf,ext=.pdf,read=.pdf,width=8.5cm]{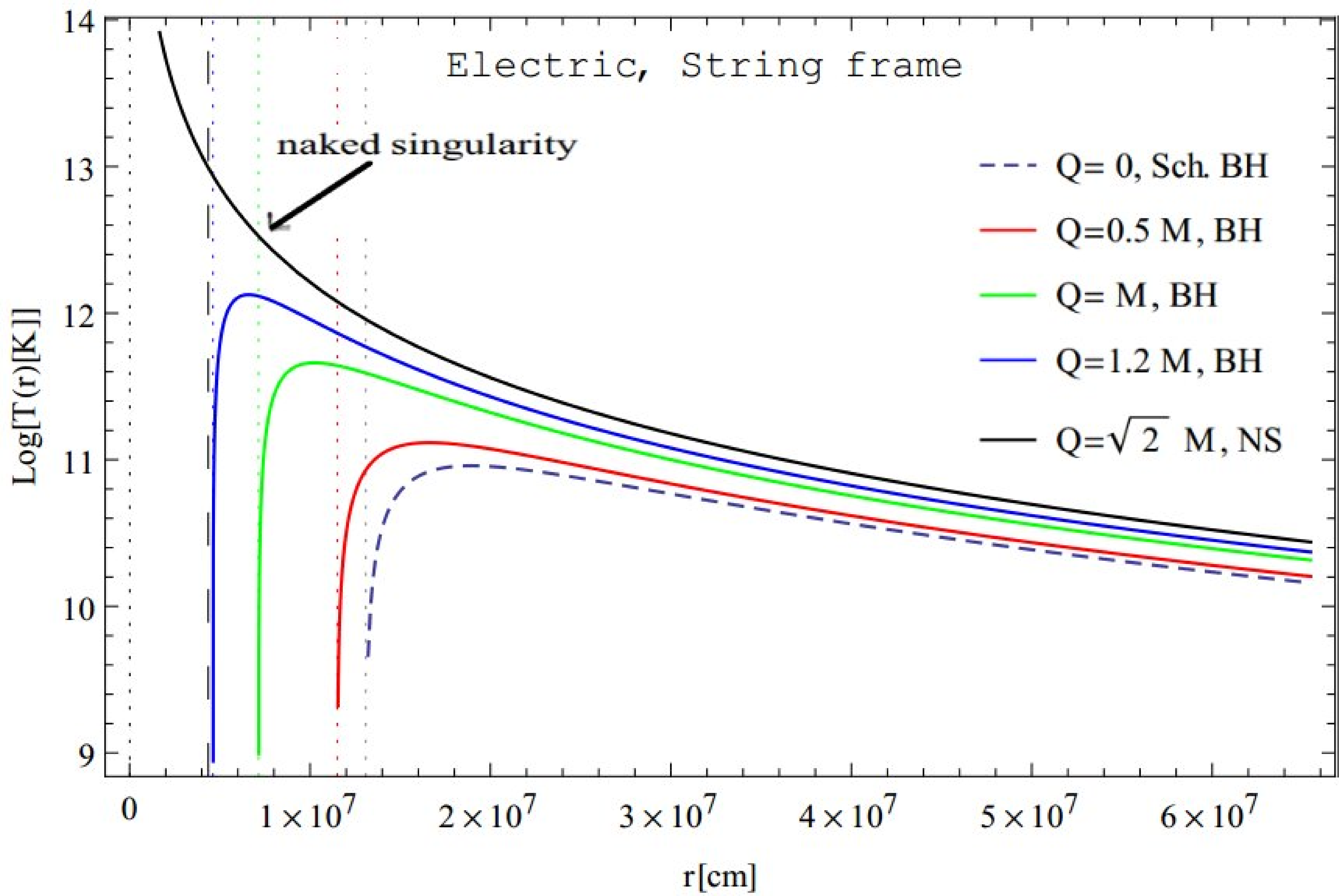}
\hspace{0.3cm}
\includegraphics[type=pdf,ext=.pdf,read=.pdf,width=8.5cm]{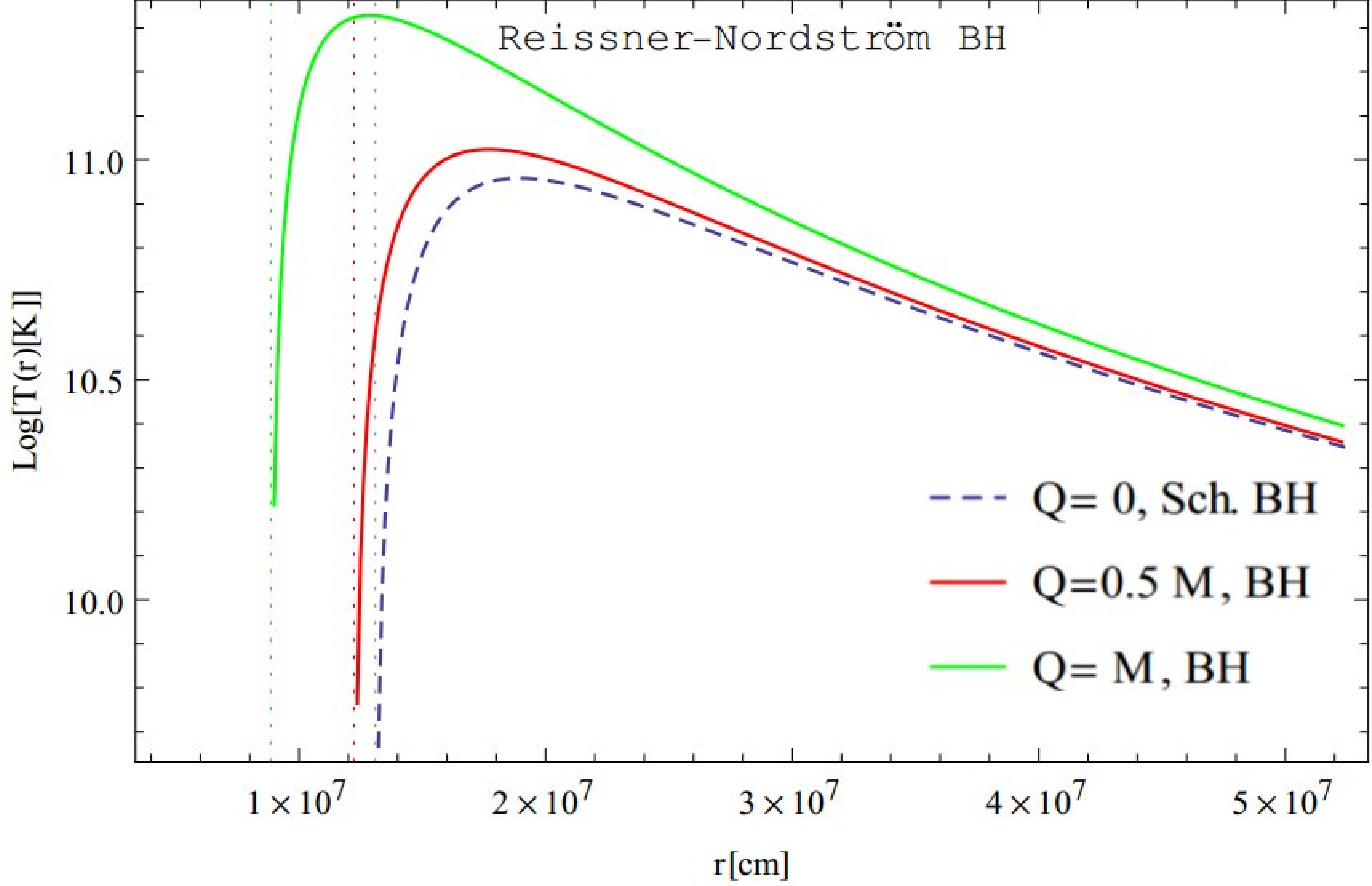}
\end{center}
\caption{Temperature distribution $T(r)$ of of the accretion disk for a GMGHS BH in EF (top left hand), magnetically charged in the SF (top right hand), electrically charged in the SF (bottom left hand) and Reissner-Nordstr\"{o}m BH (bottom right hand) plotted for different values of $Q$. We see that $T(r)\rightarrow \infty $ at NS radii $2M$ coinciding with $r_{\textmd{\scriptsize{ms}}}$ (Fig.5a), $r=r_{\textmd{\scriptsize{ms}}}\rightarrow 0$ (Fig.5c), but $T(r)$ assumes a finite value at the WH throat (Fig.5b). Fig.5d displays the behavior only for BHs. All plots except that of WH show no appreciable differences at asymptotic distances of observation.}
\label{Temp}
\end{figure*}

\begin{figure*}
\begin{center}
\includegraphics[type=pdf,ext=.pdf,read=.pdf,width=8.5cm]{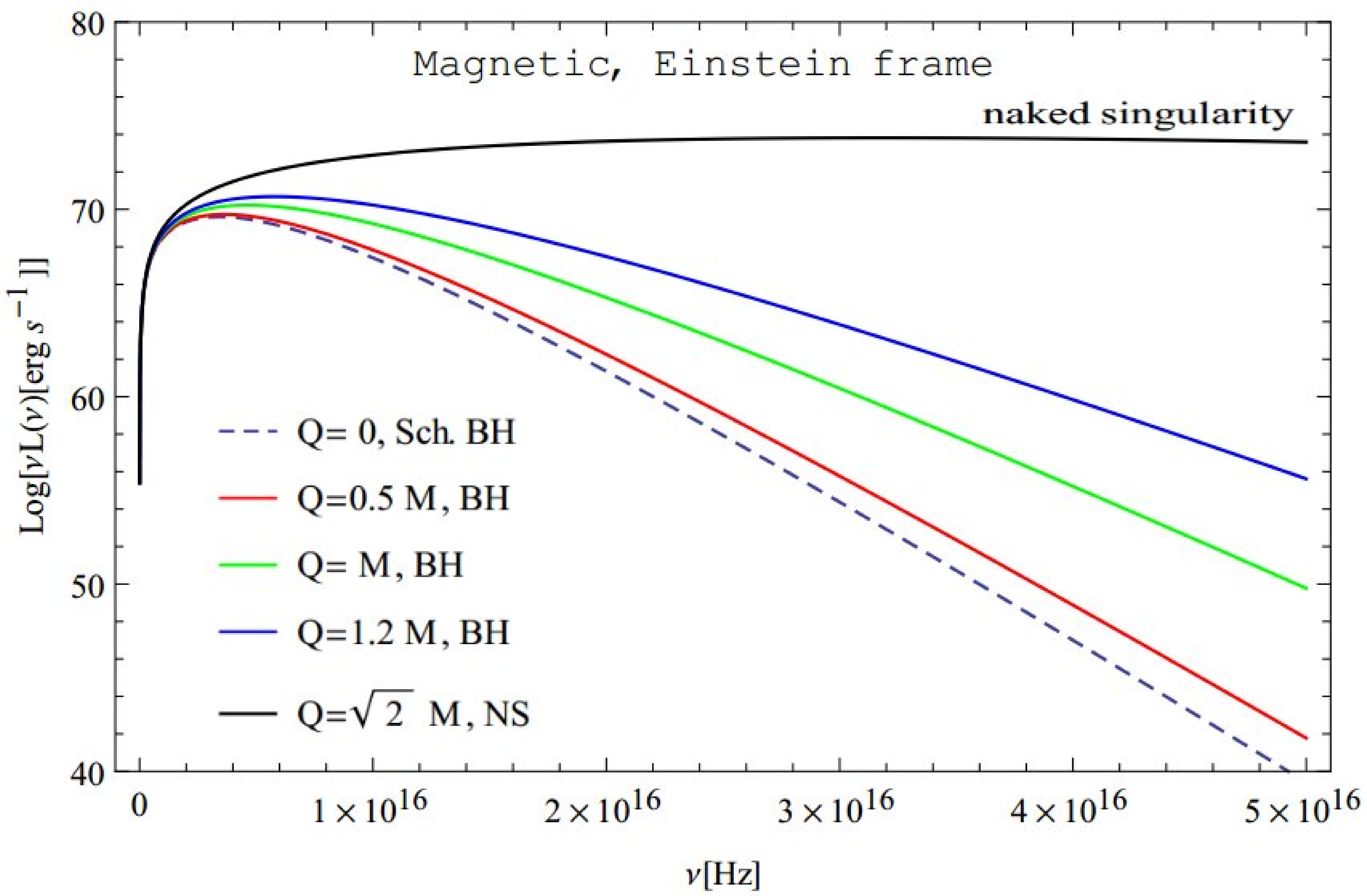}
\hspace{0.3cm}
\includegraphics[type=pdf,ext=.pdf,read=.pdf,width=8.5cm]{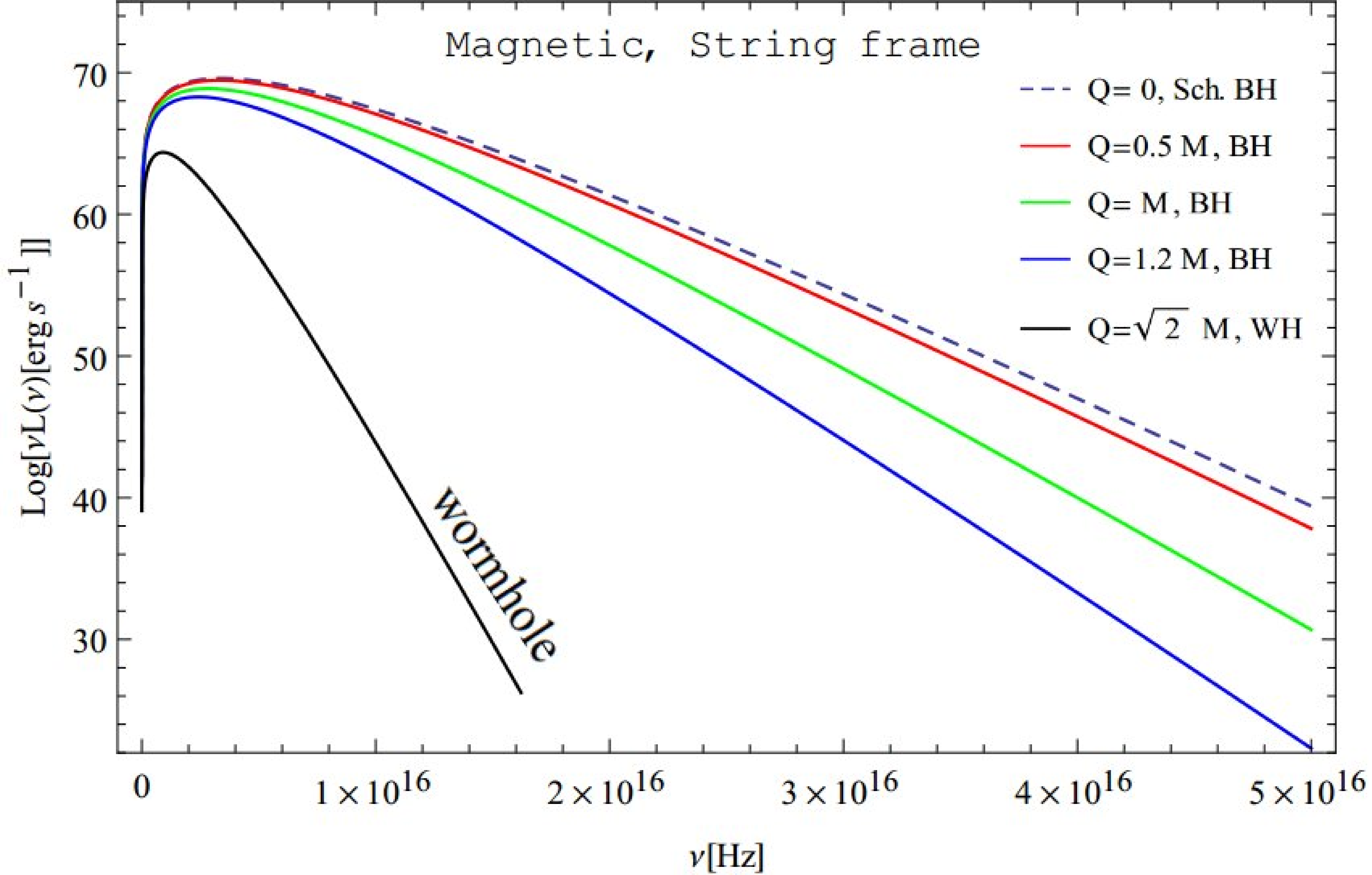} \\
\vspace{0.5cm}
\includegraphics[type=pdf,ext=.pdf,read=.pdf,width=8.5cm]{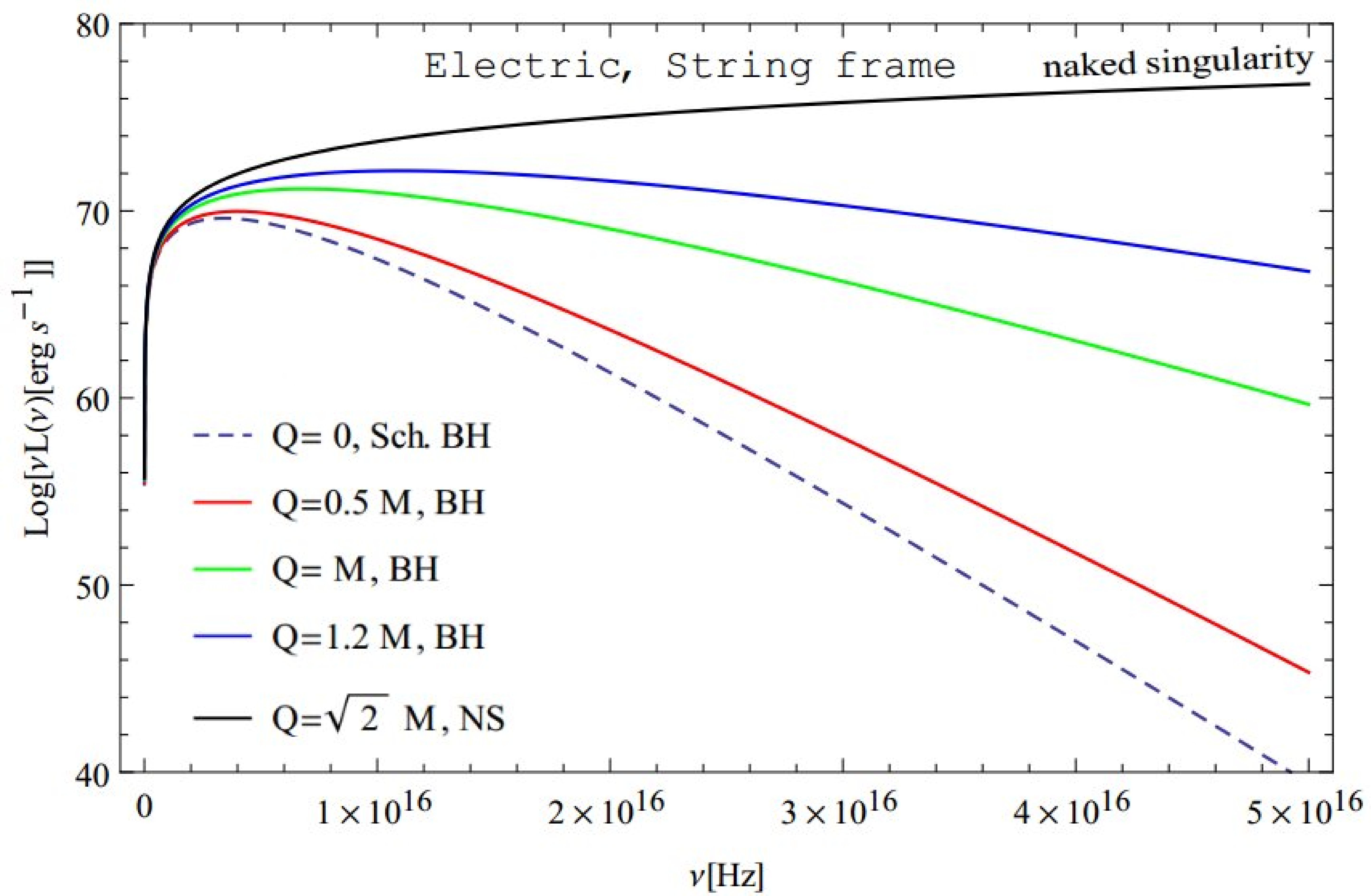}
\hspace{0.3cm}
\includegraphics[type=pdf,ext=.pdf,read=.pdf,width=8.5cm]{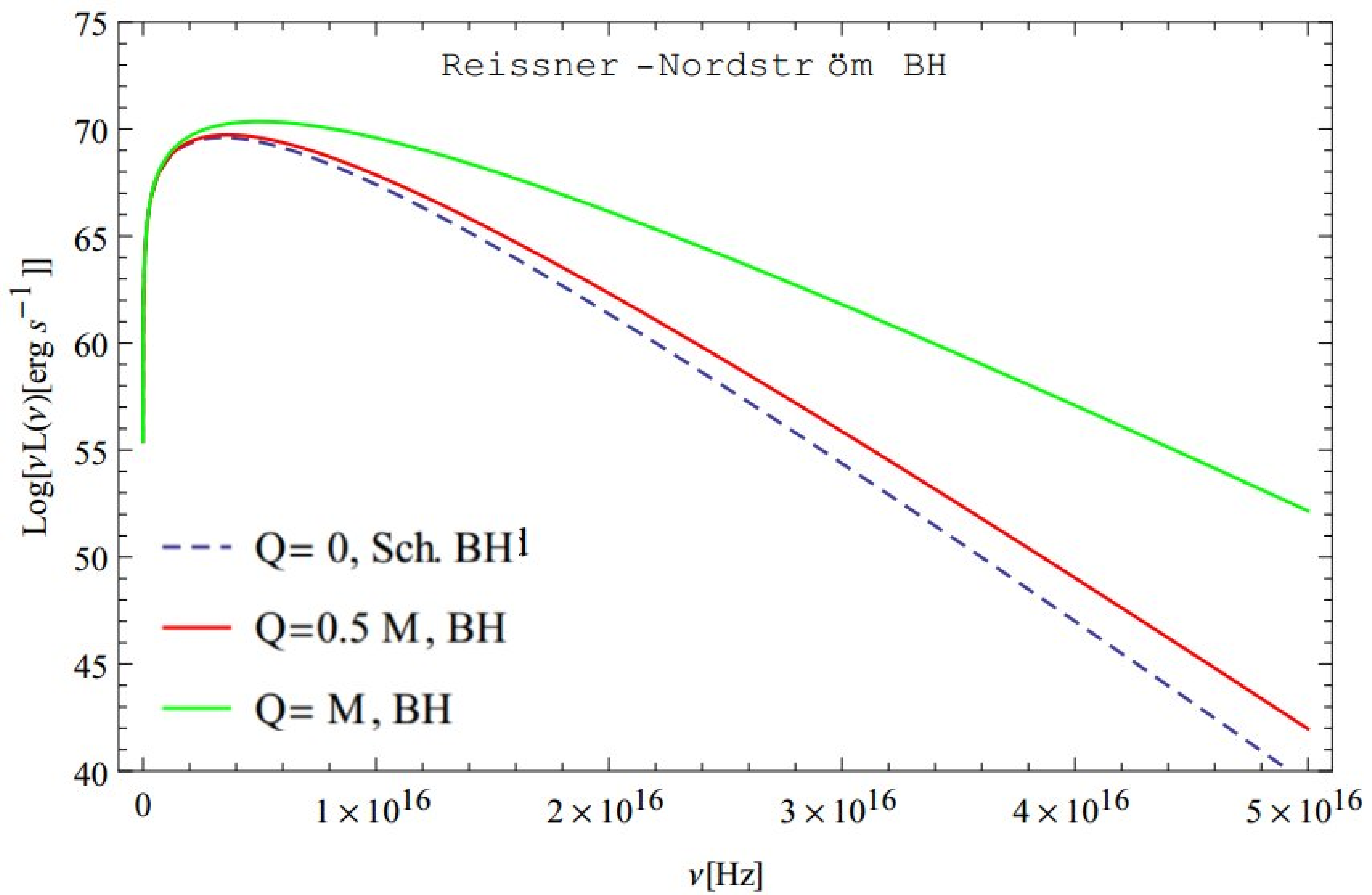}
\end{center}
\caption{The emission spectra $\protect\nu L(\protect\nu )$ vs frequency ($\protect\nu $ in Hz) of the accretion disk with inclination $i=0^{0}$ for a GMGHS BH in EF (top left hand), magnetically charged in the SF (top right hand), electrically charged in the SF (bottom left hand) and Reissner-Nordstr\"{o}m BH (bottom right hand) plotted for different values of $Q$. \ Fig. 6a shows that, at $Q\leq \protect\sqrt{2}M$, $\protect\nu L(\protect\nu )$ behaves like a BH of general relativity at the low frequency range but assumes a steady value until $\protect\nu \sim 10^{19}$ Hz, when it begins to decline to the Schwarzschild profile (not shown). Similar behavior is seen in Fig.6c. Fig.6b shows the emission spectra of the disk around WH throat. At low $\protect\nu $, the features are almost indistinguishable from those of BHs but at higher frequency, the spectra diminishes considerably. Fig.6d displays the behavior only for BHs. These features are consistent with efficiency of conversion (see summary).}
\label{Lum}
\end{figure*}

\section{Thin accretion disk}

The accretion disk is formed by particles moving in circular orbits around a
compact object, with the geodesics determined by the space-time geometry
around the object, be it a WH, BH or NS. For a static and spherically
symmetric geometry the metric is generically given by

\begin{equation}
d\tau ^{2}=-g_{tt}dt^{2}+g_{rr}dr^{2}+g_{\theta \theta }d\theta
^{2}+g_{\varphi \varphi }d\varphi ^{2}.
\end{equation}%
At and around the equator, i.e., when $\left\vert \theta -\pi /2\right\vert
\ll 1,$we assume, with Harko \textit{et al.} [27], that the metric functions
$g_{tt},g_{rr},g_{\theta \theta }$ and $g_{\varphi \varphi }$ depend only on
the radial coordinate $r$. The radial dependence of the angular velocity $%
\Omega $, of the specific energy $\widetilde{E}$, and of the specific
angular momentum $\widetilde{L}$ of particles moving in circular orbits in
the above geometry are given by
\begin{eqnarray}
&&\frac{dt}{d\tau }=\frac{\widetilde{E}}{g_{tt}}, \\
&&\frac{d\varphi }{d\tau }=\frac{\widetilde{L}}{g_{\varphi \varphi }}, \\
&&g_{tt}g_{rr}\left( \frac{dr}{d\tau }\right) ^{2}+V_{\textmd{\scriptsize{eff}}%
}\left( r\right) =\widetilde{E}^{2}.
\end{eqnarray}%
From the last equation, the effective potential $V_{\textmd{\scriptsize{eff}}}(r)$
can be obtained in the form

\begin{equation}
V_{\textmd{\scriptsize{eff}}}\left( r\right) =g_{tt}\left( 1+\frac{\widetilde{L}^{2}}{g_{\varphi \varphi }}\right).
\end{equation}%
Existence of circular orbits at any arbitrary radius $r$ in the equatorial
plane demands that $V_{\textmd{\scriptsize{eff}}}\left( r\right) =0$ and $dV_{%
\textmd{\scriptsize{eff}}}/dr=0$. These conditions allow us to write
\begin{eqnarray}
&&\widetilde{E}=\frac{g_{tt}}{\sqrt{g_{tt}-g_{\varphi \varphi }\Omega ^{2}}},
\\
&&\widetilde{L}=\frac{g_{\varphi \varphi }\Omega }{\sqrt{g_{tt}-g_{\varphi
\varphi }\Omega ^{2}}}, \\
&&\Omega =\frac{d\varphi }{dt}=\sqrt{\frac{g_{tt,r}}{g_{\varphi \varphi ,r}}}.
\end{eqnarray}

Stability of orbits depend on the signs of $d^{2}V_{\textmd{\scriptsize{eff}}}/dr^{2}$, while the condition $d^{2}V_{\textmd{\scriptsize{eff}}}/dr^{2}=0$ gives the inflection point or \textit{marginally stable} (ms) orbit or innermost stable circular orbit (ISCO) at $r=r_{\textmd{\scriptsize{ms}}}$. For the solutions under consideration, we explicitly find the $r_{\text{ms}}$ as under:

\begin{equation}
V_{\textmd{\scriptsize{eff}}}^{\textmd{\scriptsize{Mag,EF}}}= - \frac{(r-2M)^{2}(2Mr-Q^{2})}{%
rJ},
\end{equation}%
\begin{eqnarray}
V_{\textmd{\scriptsize{eff}}}^{\textmd{\scriptsize{Mag,EF}}\;\prime\prime} &=&
-\frac{4M^{3}(r-6M)}{r(Mr-Q^{2})J} \nonumber\\
&&-\frac{8M^{2}Q^{2}(3Mr-Q^{2})}{r^{3}(Mr-Q^{2})J},
\end{eqnarray}%
where
\begin{eqnarray}
J=2Mr(r-3M)-Q^{2}(r-4M). \nonumber
\end{eqnarray}
Solving $V_{\textmd{\scriptsize{eff}}}^{\textmd{\scriptsize{Mag,EF}}\;\prime\prime}=0$, we find
\begin{eqnarray}
r_{\textmd{\scriptsize{ms}}}^{\textmd{\scriptsize{Mag,EF}}}&=&2M+\left( \frac{2}{M}\right) ^{\frac{1}{3}%
}(2M^{2}-Q^{2})^{\frac{2}{3}} \nonumber \\
&&+2^{\frac{2}{3}}\{M(2M^{2}-Q^{2})\}^{\frac{1}{3}%
}.
\end{eqnarray}

Similarly, the effective potential $V_{\textmd{\scriptsize{eff}}}(r)$ and $V_{\textmd{\scriptsize{eff}}}^{\prime \prime}(r)$ for magnetic GMGHS spacetime in SF is
\begin{equation}
V_{\textmd{\scriptsize{eff}}}^{\textmd{\scriptsize{Mag,SF}}}= -\frac{2M(r-2M)^{2}}{J},
\end{equation}%
\begin{equation}
V_{\textmd{\scriptsize{eff}}}^{\textmd{\scriptsize{Mag,SF}}\;\prime\prime}=-\frac{%
2M(r-6M)(2M^{2}-Q^{2})}{r(Mr-Q^{2})J}.
\end{equation}%
Solving $V_{\textmd{\scriptsize{eff}}}^{\textmd{\scriptsize{Mag,SF}}\;\prime\prime}=0$, we find
\begin{equation}
r_{\textmd{\scriptsize{ms}}}^{\textmd{\scriptsize{Mag,SF}}}=6M.
\end{equation}

The effective potential $V_{\textmd{\scriptsize{eff}}}(r)$ and $V_{\textmd{\scriptsize{eff}}}^{\prime
\prime }(r)$ for electric GMGHS spacetime in SF
\begin{equation}
V_{\textmd{\scriptsize{eff}}}^{\textmd{\scriptsize{Elec,SF}}}=-\frac{2Mr\{M(r-2M)+Q^{2}\}^{2}}{(Q^{2}+Mr)P},
\end{equation}%
\begin{equation}
V_{\textmd{\scriptsize{eff}}}^{\textmd{\scriptsize{Elec,SF}}\;\prime\prime}=-\frac{2MN}{%
r(Mr+Q^{2})^{3}P}.
\end{equation}%
Solving $V_{\textmd{\scriptsize{eff}}}^{\textmd{\scriptsize{Elec,SF}}\;\prime\prime}=0$, we find the
marginally stable orbit
\begin{eqnarray}
r_{\textmd{\scriptsize{ms}}}^{\textmd{\scriptsize{Elec,SF}}}&=&\frac{2M^{2}-Q^{2}}{M}+2^{\frac{4}{3}%
}M^{2}(2M^{2}-Q^{2})K^{-\frac{1}{3}} \nonumber \\
&&+\frac{2^{\frac{2}{3}}K^{\frac{1}{3}}}{%
2M^{2}+Q^{2}},
\end{eqnarray}%
where
\begin{eqnarray}
P &=&2M^{2}r(r-3M)-Q^{2}(2M^{2}-3Mr-Q^{2}), \\
N &=&2M^{5}r^{2}(r-6M)+M^{3}Q^{2}r^{3}+M^{2}Q^{4}\left(4M^{2} \right.  \nonumber \\
&& \left.-3Mr+3r^{2}\right)-Q^{6}(4M^{2}-3Mr-Q^{2}), \\
K &=&16M^{9}-8M^{5}Q^{4}+MQ^{8}+MQ^{2}(Q^{4}-4M^{4})^{\frac{3}{2}}.
\end{eqnarray}

We assume geometrically thin accretion disk, which means that the disk height $H$
above the equator is much smaller than the characteristic radius $R$ of the
disk, $H\ll R$. The disk is assumed to be in hydrodynamical equilibrium
stabilizing its vertical size, with the pressure and vertical entropy
gradient being negligible. An efficient cooling mechanism via heat loss by
radiation over the disk surface is assumed to be functioning in the disk,
which prevents the disk from collecting the heat generated by stresses and
dynamical friction. The thin disk has an inner edge defined by the \textit{%
marginally stable} circular radius $r_{\textmd{\scriptsize{ms}}}$, while the
orbits at higher radii are Keplerian.

In the above approximation, Page and Thorne [24], using the rest mass
conservation law, showed that the time averaged rate of rest mass accretion $%
dM_{0}/dt$ is independent of the radius: $\dot{M}_{0}\equiv dM_{0}/dt=-2\pi
ru^{r}\Sigma =$ \textmd{const}. (Here $t$ and $r$ are the coordinate time
and radial coordinates respectively, $u^{r}$ is the radial component of the
four velocity $u^{\mu }$ of the accreting particles and $\Sigma $ is the
averaged surface density of the disk). In the steady-state thin disk model,
the orbiting particles have $\Omega $ , $\widetilde{E}$ and $\widetilde{L}$
that depend only on the radii of the orbits. We omit other technical details
here (see [27]), but quote only the relevant formulas below.

The flux $F$ of the radiant energy over the disk can be expressed in terms
of $\Omega $ , $\widetilde{E}$ and $\widetilde{L}$ as [22-24]

\begin{equation}
F(r)= -\frac{\dot{M}_{0}}{4\pi \sqrt{-g}}\frac{\Omega _{,r}}{%
\left( \widetilde{E}-\Omega \widetilde{L}\right) ^{2}}\int_{r_{\textmd{\scriptsize{ms}}%
}}^{r}\left( \widetilde{E}-\Omega \widetilde{L}\right) \widetilde{L}_{,r}dr.
\end{equation}
The disc is supposed to be in thermodynamical equilibrium, so the radiation
flux emitted by the disk surface will follow Stefan-Boltzmann law:%
\begin{equation}
F\left( r\right) =\sigma T^{4}\left( r\right) ,
\end{equation}%
where $\sigma $ is the Stefan-Boltzmann constant. The observed luminosity $%
L\left( \nu \right) $ has a redshifted black body spectrum [36]

$$L_{\nu }=4\pi \textmd{d}^{2}I(\nu )=\frac{8\pi h\cos {i}}{c^{2}}\int_{r_{\textmd{\scriptsize{in}}}}^{r_{\textmd{\scriptsize{f}}}}\int_{0}^{2\pi }\frac{\nu _{e}^{3}rdrd\varphi }{\textmd{Exp}\left[ \frac{h\nu _{e}}{k_{B}T}\right] -1},$$
where $i$ is the disk inclination angle, d is the distance between the
observer and the center of the disk, $r_{\textmd{\scriptsize{in}}}$ and $r_{\textmd{\scriptsize{f}}}$ are
the inner and outer radii of the disc, $h$ is the Planck constant, $\nu _{e}$
is the emission frequency, $I(\nu )$ is the Planck distribution, and $k_{B}$
is the Boltzmann constant. The observed photons are redshifted and their
frequency $\nu $ is related to the emitted ones in the following way $\nu
_{e}=(1+z)\nu $. The redshift factor $(1+z)$ has the form [27]:

\begin{equation}
(1+z)=\frac{1+\Omega r\sin {\varphi }\sin {i}}{\sqrt{g_{tt}-\Omega
^{2}g_{\varphi \varphi }}},
\end{equation}%
where the light bending effect is neglected.

Another important characteristic of the accretion disk is its efficiency $%
\epsilon$, which quantifies the ability with which the central body
converts the accreting mass into radiation. The efficiency is measured at
infinity and it is defined as the ratio of two rates: the rate of energy of
the photons emitted from the disk surface and the rate with which the
mass-energy is transported to the central body. If all photons reach
infinity, an estimate of the efficiency is given by the specific energy of
the accreting particles measured at the marginally stable orbit [25]:

\begin{equation}
\epsilon =1-\widetilde{E}\left( r_{\textmd{\scriptsize{ms}}}\right) .
\end{equation}%
The Eqs.(18-20,33-36) are valid for any static spherically symmetric spacetime and
hence valid both in the EF and SF since spherical symmetry is preserved
under conformal transfomation.

\section{GMGHS solutions: kinematic and accretion features}

We shall consider for illustration a central compact object of mass $%
M=15M_{\odot }$ with an accretion rate $\dot{M_{0}}=10^{18}$gm.sec$^{-1}$
and assume that it possesses magnetic charge $Q$ with the spacetime
described by stringy BH solutions (2), (8) and (11).

We examine how kinematic features of the concerned solutions differ from
those of Schwarzschild BH. First, the plot for $V_{\textmd{\scriptsize{eff}}}$ are shown in
Figs.1a-1c, for $r\in \lbrack r_{\textmd{\scriptsize{eh}}}$,$\infty )$, while the parameter
$Q$ is taken to assume values $0.5M$, $M$, $1.2M$ and $\sqrt{2}M$. The plots
reveal that all potentials show finite maxima for $Q^{2}<2M^{2}$ but diverge
at $r_{\textmd{\scriptsize{eh}}}^{\textmd{\scriptsize{Mag,EF}}}=r_{\textmd{\scriptsize{eh}}}^{%
\textmd{\scriptsize{Mag,SF}}}=2M$ in the extreme case $Q^{2}=2M^{2}$ and at $r_{%
\textmd{\scriptsize{eh}}}^{\textmd{\scriptsize{Elec,SF}}}=0$ corresponding to a NS, see
after Eq.(12). The effective potentials of the Schwarzschild and
Reissner-Nordstr\"{o}m solution are plotted in Fig.1d for comparison. An
interesting feature is that all the potentials in the EF and SF show smooth
asymptotic fall-off approaching the Schwarzschild curve from above. In
contrast, the potential $V_{\textmd{\scriptsize{eff}}}^{\textmd{\scriptsize{Elec,SF}}}$, after reaching a
local maximum at $r_{\textmd{\scriptsize{ms}}}$, dips below the Schwarzschild curve
intersecting it at a radius given by (Fig.1c):
\begin{equation}
r|_{V_{\textmd{\scriptsize{eff}}}^{\textmd{\scriptsize{Elec,SF}}}=V_{\textmd{\scriptsize{eff}}}^{\textmd{\scriptsize{Sch}}}}=\frac{%
4M^{2}-Q^{2}+\sqrt{16M^{4}+Q^{4}}}{2M}.
\end{equation}

Second, we can observe similar features in Figs.2a-2d demonstrating the
effect of $Q$ on specific angular momenta $\widetilde{L}$. Note that, in all
these plots, the distance scale is $r\in \lbrack r_{\textmd{\scriptsize{eh}}}$,$\infty )$
except for $Q^{2}=2M^{2}$, when $r_{\textmd{\scriptsize{eh}}}=0$ (singularity). However, in
case of $\widetilde{L}^{\textmd{\scriptsize{Elec,SF}}}$, its intersection point
with that of the Schwarzschild BH differs from the intersection point of
effective potential and is given by
\begin{equation}
r|_{\widetilde{L}^{\textmd{\scriptsize{Elec,SF}}}=\widetilde{L}^{\textmd{\scriptsize{Sch}%
}}}=\frac{8M^{2}-Q^{2}+\sqrt{32M^{4}+Q^{4}}}{2M}.
\end{equation}

Third, Figs.3a-3d show influence of $Q$ on the specific orbital energies.
Fig.3a shows that energies of a particle orbiting a Schwarzschild BH is
higher than that orbiting a magnetic\ GMGHS BH at the same radius in the EF.
A similar behavior is seen for orbits in the electric\ GMGHS BH in the SF
(Fig.3c). This behavior is similar to that in the Reissner-Nordstr\"{o}m
case (Fig.3d). However, exactly the opposite behavior is shown by the orbits
in the magnetic\ GMGHS BH in the SF, where orbital energies for particles in
the Schwarzschild BH spacetime are at the lowest (Fig.3b).

Finally, we focus on the observable emissivity features such as the
radiation flux, luminosity, temperature and conversion efficiency. Figs.4a-d
display the flux of radiation $F\left( r\right) $ emitted by the disk
between $r_{\textmd{\scriptsize{ms}}}$ and received at an arbitrary radius $r$ [Eq.(33)],
Figs 5a-d show variation of temperature over the disk from $r_{\textmd{\scriptsize{ms}}}$
to an arbitrary radius and Figs.6a-d show observed luminosity variations on
a logarithmic scale over different frequency ranges. The conversion
efficiency $\epsilon $ of the accreting mass into radiation, measured at
infinity is given by Eq.(36). In Tab.1, we show the radiation properties of
the accretion disks using marginally stable orbits $r_{\textmd{\scriptsize{ms}}}$ and $%
\epsilon $ with the parameter $Q$ in the range used in the previous plots.

\begin{table*}[!ht]
\caption{The $r_{\textmd{\scriptsize{ms}}}$ and the efficiency $\epsilon $ for GMGHS and Reissner-Nordstr\"{o}m BHs, all having a mass $M=15M_{\odot }$
         with the accretion rate $\dot{M}_{0}=10^{18}$ gm.sec$^{-1}$. The general relativistic Schwarzschild BH corresponds to $Q=0$.}
  \begin{tabular}{|c|c|c|c|c|c|c|c|c|}
  \hline
  $Q$ & \multicolumn{2}{c|}{Magnetic, EF} & \multicolumn{2}{c|}{Magnetic, SF}
  & \multicolumn{2}{c|}{Electric, SF} & \multicolumn{2}{c|}{Reissner-Nordstr\"{o}m BH} \\
  \cline{2-9}
     & $r_{\textmd{\scriptsize{ms}}}$ [$M$] & $\epsilon $ & $r_{\textmd{\scriptsize{ms}}}$ [$M$] & $\epsilon $
  & $r_{\textmd{\scriptsize{ms}}}$ [$M$] & $\epsilon $ & $r_{\textmd{\scriptsize{ms}}}$ [$M$] & $\epsilon $ \\
  \hline
  & \multicolumn{8}{c|}{BH} \\
  $0$ (Sch) & 6.0000 & 0.0572 & \multicolumn{6}{c|}{} \\
  $0.5M$ & 5.7426 & 0.0607 & 6.0000 & 0.0506 & 5.2746 & 0.0715 & 5.6066 &
  0.0608 (BH) \\
  $M$ & 4.8473 & 0.0768 & 6.0000 & 0.0298 & 3.2743 & 0.1361 & 4.0000 & 0.0814
  (extr. BH) \\
  $1.2M$ & 4.1644 & 0.0950 & 6.0000 & 0.0170 & 2.1163 & 0.2080 & 2.4548 &
  0.1149 (NS) \\
  - & \multicolumn{2}{c|}{NS} & \multicolumn{2}{c|}{WH}
  & \multicolumn{2}{c|}{NS} &  &  \\
  $\sqrt{2}M$ & 2.0000 & 1.0000 & 6.0000 & 0 & 0 & 1.0000 & 2.3129 & 0.0773
  (NS) \\ \hline
  \end{tabular}
\end{table*}

Table 1 demonstrates the variation in the location of the inner disk edge
with the changing charge $Q$. For GMGHS BH in EF, we notice that the higher
values of charge $Q$ are, the closer are the marginally stable orbits to the
center. However, for magnetically charged GMGHS BH in SF, we see that its
conversion efficiency for $Q\sim 0.3M$ mimics that of the Schwarzschild BH,
both being $0.0572.$ For electrically charged GMGHS BH in SF, the efficiency
increases to over $20\%$, whereas, interestingly, that for magnetically
charged GMGHS BH in SF the efficiency decreases to lower than $2.45\%$.
These features are characteristic of the frames chosen for describing BHs.

\begin{table*}[!ht]
 \caption{Comparison with BH of luminosity spectra from accretion disk around
  different extreme GMGHS central objects.}
 \begin{tabular}{|c|c|c|c|c|}
  \hline
  $\nu$[1/s] & \multicolumn{4}{c|}{$\nu L(\nu)$[erg s$^{-1}$]} \\
  \cline{2-5}
            & \multicolumn{2}{c|}{Naked singularity} & Wormhole & Schwarzschild \\
  \cline{2-4}
             & Magnetic, EF    & Electric, SF          & Magnetic, SF & black hole\\
  \hline
  $10^{12}$  & $1.15\times 10^{24}$   & $1.63\times 10^{24}$  & $8.75\times 10^{22}$   & $1.15\times 10^{24}$   \\
  $10^{13}$  & $1.13\times 10^{26}$   & $1.60\times 10^{26}$  & $8.20\times 10^{24}$   & $1.11\times 10^{26}$   \\
  $10^{14}$  & $1.07\times 10^{28}$   & $1.51\times 10^{28}$  & $6.21\times 10^{26}$   & $9.84\times 10^{27}$   \\
  $10^{15}$  & $9.08\times 10^{29}$   & $1.31\times 10^{30}$  & $8.95\times 10^{27}$   & $5.57\times 10^{29}$   \\
  $10^{16}$  & $4.57\times 10^{31}$   & $1.02\times 10^{32}$  & $1.13\times 10^{19}$   & $1.90\times 10^{29}$   \\
  $10^{17}$  & $2.13\times 10^{31}$   & $8.43\times 10^{33}$  & $\sim 0$               & $1.01\times 10^{0}$    \\
  $10^{18}$  & $3.34\times 10^{15}$   & $6.77\times 10^{35}$  & $\sim 0$               & $\sim 0$               \\
  $10^{19}$  & $\sim 0$               & $8.31\times 10^{36}$  & $\sim 0$               & $\sim 0$               \\
  $10^{20}$  & $\sim 0$               & $1.10\times 10^{27}$  & $\sim 0$               & $\sim 0$               \\
  $10^{21}$  & $\sim 0$               & $\sim 0$              & $\sim 0$               & $\sim 0$               \\
  \hline
\end{tabular}
\end{table*}

The Figs.1-6 show different kinematic and emissivity parameters for
different values of $Q$ related to dilatonic charge $D$ [Eq.(5)]. We have
assumed the central mass to be $M=15M_{\odot },$ and mass accretion rate $%
\dot{M}_{0}=10^{18}$ gm$.$sec$^{-1}$.

\section{Summary}

The spacetime structure of GMGHS BHs differ from that of the Schwarzschild BH in many important ways, which are expected to show up in their kinematic and accretion disk properties. In the present paper, we analyzed thin accretion disk properties around magnetically and electrically charged GMGHS BHs in EF and in SF. The physical parameters describing the disk such as the effective potential, radiation flux, temperature and emissivity profiles have been explicitly obtained for several values of the parameter $Q$, that in turn correspond to several values for dilationic charge $D$ for a given mass $M$. All the astrophysical quantities related to the observable properties of the accretion disk in the two frames have been compared with those for the Schwarzschild BH of the same mass. We considered as a toy model a stellar sized compact object of mass $M=15M_{\odot}$ with an accretion rate $\dot{M}_{0}=10^{18}$ gm.sec$^{-1}$ and assumed that its accretion properties can be described by the those of the GMGHS spacetimes including their extreme limits of NS and WHs. Our aim has been to examine whether the kinematic and emissivity properties significantly change when the central object changes.

The main conclusions of our analyses are as follows: Kinematic properties were already analyzed in Sec.4 with corresponding figures and need not be repeated here. Suffice it to say that at the NS radius, all kinematic quantities diverge, as expected. The emissivity properties of GMGHS BHs do not appreciably differ from those of the Schwarzschild or Reissner-Nordstr\"{o}m BH for $Q<\sqrt{2}M$. This conclusion is in accord with the strong lensing properties of GMGHS BH studied by Bhadra [6]. However, in the extreme limit $Q=\sqrt{2}M$, the GMGHS BH in the EF yields NS. In the SF, there are two GMGHS solutions, one is electrically charged and the other is magnetically charged. In the extreme limit the former yields NS and the latter yields WHs. These latter types of geometries are interesting in their own right since the existence of NS is associated with the no-hair theorem and a recent work includes accretion properties in the JNW NS [42,43]. Also, the horizonless WHs are seriously considered as candidates for mimicking initial post-merger ring down signals characteristic of BH horizon [55-60]. While the emissivity properties of ordinary GMGHS BHs in EF and in SF do not differ appreciably from those of the Schwarzschild BH, their extreme counterparts studied here show that they differ quite significantly from those of the Schwarzschild BH. These differences provide yet another avenue to distinguish between ordinary and extremal objects.

A very interesting result, qualitatively similar to the one obtained by Torres [36] for boson stars, is presented in Table 2 and correspondingly in Figs.6a-d for our toy model. The table shows the difference with Schwarzschild BH, when NS and WH are concerned. From these, we see that for BHs ($Q<\sqrt{2}M$), and WHs in the SF ($Q=\sqrt{2}M$) the spectra decay rapidly for $\nu >10^{16}$ Hz but for NS ($Q=\sqrt{2}M$), the luminosity (Figs. 6a,6c) does not decay until a frequency $\nu >10^{18}$ Hz (for magnetic NS in EF) beyond which it becomes nearly invisible. The same thing happens for electric NS in SF beyond $\nu>10^{20}$ Hz. In either case, there is almost an \textit{infinite} increase in observed luminosity compared to nearly invisible WH and BH at the same frequency. Thus NS should be the brightest objects in the sky (like QSOs). This conclusion is supported by the efficiency $\epsilon =1$ for NS in Table 1. It thus seems that, at the singular radius $r=r_{\textmd{\scriptsize{ms}}}=2M$ (first column of the Table, see Figs. 6a,c), all infalling matter is minced into radiation all of which then escape to us.

It would be of interest to examine the physical reason as to why such rapid decays in the luminosity spectra occur at higher frequencies but it is evident that observable distinctions exist among different objects described by GMGHS. Understandably, a more adequate and realistic model for the central object should include spin. A choice for this purpose could be the spinning Kerr-Sen solution [61] but it is expected that spin might not drastically alter the main conclusions derived here. We keep it as an open problem for future work.


\begin{acknowledgments}
We thank an anonymous referee for useful suggestions. The reported study was funded by RFBR according to the research project No. 18-32-00377.
\end{acknowledgments}


\end{document}